\documentclass[12pt]{article}
\usepackage{booktabs,tabularx}

\usepackage{algorithm}
\usepackage{enumerate}
\usepackage{algpseudocode}
\usepackage{algorithm}
\usepackage{algpseudocode}
\usepackage{fullpage}
\usepackage{amssymb,amsmath}
\usepackage{amsfonts}
\usepackage{amsthm,bm}
\usepackage{graphicx}
\usepackage[dvipdfm]{geometry}
\usepackage{times}
\usepackage{natbib}

\usepackage{multirow}
\usepackage{color}
\usepackage{ulem}
\usepackage{graphicx,amsmath}
\usepackage{lettrine}
\usepackage{dcolumn}

\usepackage{rotating, graphicx}

\usepackage{setspace}
\usepackage{mathtools}

\usepackage{soul}
\setstcolor{red}

\usepackage{url}
\makeatletter
\def\url@urlstyle{%
  \@ifundefined{selectfont}{\def\UrlFont{\sf}}{\def\UrlFont{\small\ttfamily}}}
\makeatother 

\begin{document}
\title{How Search Engine Advertising Affects Sales over Time: An Empirical Investigation}
\author{Yanwu Yang$^1$, Kang Zhao$^2$, Daniel Zeng$^3$ and Bernard Jim Jansen$^4$\\
\footnotesize{$^1$School of Management, Huazhong University of Science and Technology, Wuhan 430074, China}\\
\footnotesize{$^2$Department of Business Analytics, University of Iowa, Iowa City, IA 52242}\\
\footnotesize{$^3$Department of Management Information Systems, University of Arizona, Tucson, AZ 85721}\\
\footnotesize{$^4$College of Information Sciences and Technology, The Pennsylvania State University, University Park, PA 16802}\\
\footnotesize{yangyanwu.isec@gmail.com}} 
\date{}
\maketitle

\begin{abstract}
  As a mainstream marketing channel on the Internet, Search Engine Advertising (SEA) has a huge business impact and attracts a plethora of attention from both academia and industry.
  One important goal of advertising is to increase sales. Nevertheless, while previous research has studied multiple factors that are potentially related to the outcome of SEA campaigns, effects of these factors on actual sales generated by SEA remain understudied. It is also unclear whether and how such effects change over time in highly dynamic SEA campaigns. As the first empirical investigation of the dynamic advertisement-sales relationship in SEA, this study builds an advertising response model within a time-varying coefficient (TVC) modeling framework, and estimates the model using a unique dataset from a large E-Commerce retailer in the United States. Results reveal the effects of the advertising expenditure, consumer behaviors and advertisement characteristics on realized sales, and demonstrate that such effects on sales do change over time in non-linear ways. More importantly, we find that carryover has a stronger effect in generating sales than direct response does, conversion rate is much more important than click-through rate, and ad position does not have significant effects on sales. These findings have direct implications for advertisers to launch more effective SEA campaigns. 

\end{abstract}

\textbf{Keywords:} Search engine advertising; Ad-sales relationship; Electronic commerce; Advertising analytics; Business intelligence.

\section{Introduction}
During the past decade, search engine advertising (SEA) has become one of the most prominent outlets for online advertising. Through SEA, advertisers pay search engines to display their advertisements related to search queries along with organic results on search engine result pages (SERPs). The economic impact of SEA has been well documented. According to the Interactive Advertising Bureau (2018), the revenue for SEA in the U.S.\ alone exceeded 22 billions USD during the first half of 2018, accounting for nearly half of the total revenue for online advertising during that period. With the high expenditure on SEA, advertisers are eager to know what factors drive the outcome of SEA campaigns. Although SEA success can be measured in different ways (e.g., online traffic and brand awareness), sales are typically one of the most important performance criteria  advertisers care about, especially in E-Commerce \citep{Sun2020}. Indeed, a better understanding of the ad-sales relationship can help advertisers make more effective investment decisions in SEA campaigns. 



SEA is a much more dynamic and evolving market \citep{YaoMela2011} than traditional marketing channels (e.g., newspapers and TV). 
At the core of SEA is real-time position auctions run by search engines to determine which ads to be displayed on a SERP and their rankings. As participants of these auctions, advertisers need to make decisions on expenditures by considering a range of factors related to consumer behaviors (e.g., ad clicks and product purchases), characteristics of advertisement (e.g., ad positions) as well as competitions from other advertisers. Note that the values of these factors to advertisers could change over time \citep{GoogleAdwords2019, Comscore2008}. As a result, it has been well-recognized by advertisers that strategies governing SEA campaigns need to be dynamically adjusted in order to achieve more sales \citep{DaSilva2018, Adaplo2019, George2019, ZhangFeng2011, Ye2014, Yang2015}. 

Although business needs of understanding the dynamic ad-sale relationship in SEA are clear and important, very little research has investigated the drivers of sales generated from SEA. Previous studies of SEA have mainly investigated measures related to consumers' clicks on ads. While intuitive and easy to obtain, clicks on ads are not equal to sales. Even converted consumers from a same ad may purchase the advertised product in small or large quantities, leading to different amount of sales. Although an earlier study attempted to associate the advertising expenditure with sales \citep{Blake2015}, the model did not consider other important factors beyond the expenditure, and analyze the ad-sales relationship in a static way. 

Therefore, a critical gap exists for researchers to investigate the effects of a comprehensive set of factors on sales from SEA over time. For example, SEA has been commonly recognized as a form of direct response advertising and a short-term investment \citep{Pabich2011}, because advertisers pay to attract traffics to their websites to generate online transactions immediately. Many advertisers assume such effects to stop after the expenditure on SEA campaigns stops. However, while researchers have questioned such an assumption \citep{Challis2014, Baadsgaard2017, Membrillo2018}, there lacks an empirical evidence on whether the effectiveness of SEA occurs in a direct (i.e., immediate) or indirect (i.e., time-lagged) manner. Besides the expenditure, other factors related to consumer behaviors (e.g., click-through rate (CTR), conversion rate (CVR), cost-per-click (CPC)), and advertisement characteristics (e.g., ad positions in SERPs) could also affect sales. Nevertheless, no studies have directly compared their effects with each other or analyzed how their effects change over time.

A better understanding of the dynamic ad-sales relationship in SEA can have tremendous values. 
Because spending more in SEA does not necessarily lead to higher sales \citep{Fischer2011, Yang2015}, advertisers need to base their budget allocation on advertising dynamics in SEA to maximize their returns in the ever changing market. Such real-time decision support in SEA is especially important for advertisers from small and medium enterprises, who represent the main revenue sources for search engines but have limited resources to understand such complexity and optimize their budgets for their SEA campaigns \citep{Anderson2005} .
To address the aforementioned gap and challenges, this paper represents the first effort to empirically explore the dynamic advertising-sales relationship in SEA. 
Specifically, this research builds an advertising response model within a time-varying coefficient (TVC) modeling framework \citep{Naik2015} to capture the dynamic nature of SEA markets. We choose the partial adjustment model \citep{Vanhonacker1983, Kohler2016} to examine carryover effects in SEA. We empirically estimate our response model using a unique panel dataset collected from SEA campaigns by a large U.S. E-Commerce retailer.

Our study has several interesting findings with important managerial and theoretical implications. \textit{First}, we reveal for the first time the dynamic nature of SEA: the effects of various factors on SEA sales do change over time. 
This implies that advertisers must continuously track, predict and adjust their advertising strategies based on the real-time effectiveness of their SEA campaigns. 

\textit{Second}, the ad-sales relationship in SEA demonstrates strong carryover effects. This contradicts the commonly-held view that investment in SEA can generate sales immediately. Instead, SEA advertiser may need to be more patient and willing to commit longer-term efforts. Meanwhile, without considering carryover effects, the immediate effect of the advertising expenditure on sales may be overestimated.

\textit{Third}, among measures of consumer behaviors, CVR plays a much more important role than CTR in predicting sales through SEA. This highlights the value of CVR in increasing sales, although major search engines often encourage advertisers to focus on improving CTR in the current pay-per-click (PPC) scheme. 

\textit{Last}, this research also reveals that ad position has no significant effect on sales. In other words, instead of always bidding for the highest ad position on SERPs, a more cost-effective way for SEA advertisers is to get their ads displayed at lower ad positions. 

The remainder of this paper is organized as follows. Section 2 presents a brief survey of related research. 
This is followed by descriptions of our data and key variables used in this research in Section 3. In Section 4, we discuss basic principles of the time-varying modeling framework and present a time-varying response model for SEA. Empirical results are listed in Section 5. The last section concludes with managerial implications and theoretical contributions, and future research directions.

\section{Related Literature and Theoretical Background}

This paper studies the dynamic ad-sales relationship in SEA and is related to literature from three streams of research: (i) factors related to the performance of SEA campaigns, (ii) dynamic processes and advertising decisions in SEA, and (iii) time-varying modeling. 

The growth of SEA has motivated studies that investigated factors for the success of SEA campaigns. An advertiser's expenditure is critical for its SEA campaign to gain more visibility and revenue. Meanwhile, there are two possible ways the advertising expenditure affects sales: (1) Direct via immediate responses (a.k.a., short-term advertising elasticity)--the current advertising expenditure affects current sales directly and immediately \citep{Assmus1984,Sethuraman2011}; or (2) Indirect via carryover effects--a considerable time lag exists between the display of an advertisement and sales of the advertised product. In other words, a certain amount of sales generated by an advertisement is not achieved immediately after the expenditure and deployment of the advertisement. Previous research has reported empirical evidence of carryover effects in online advertising channels. For example, \citet{Johnson2017} analyzed 432 online display advertising field experiments on the Google Display Network, and found most campaigns have a modest and positive carryover. \citet{Archak2012} demonstrated that considering carryover effects can better explain revenues in SEA markets. However, an empirical evidence on carryover effects in SEA is still lacking.

Besides the expenditure, other well-recognized factors for SEA success include ad positions and consumer behaviors over ads. In SEA, ad position has been a key factor that advertisers compete for \citep{Agarwal2011, Jansen2013}.
According to the theory of serial position effects \citep{Ebbinghaus1913}, the position of an ad impacts user's perception and interaction behaviors during a web search session \citep{Jansen2013}. Generally, higher ad positions are expected to generate more traffics and sales for advertisers. Therefore, in Google Adwords, higher ad positions are awarded to higher bidders if competing ads have the same relevance and quality. 

While advertisers care about how many times their ads are displayed to consumers, it is more important how consumers interact with their ads. Therefore, major search engines predominantly adopt the pay-per-click scheme and charge an advertiser only when their ads are clicked by consumers. There are three important measures of consumer behaviors in the PPC scheme: \textit{Click-through-rate} is the ratio of clicks on an ad over impressions by consumers. \textit{Conversion rate} is the ratio of conversions (e.g., making a purchase) over the total number of ad clicks. In fact, many previous studies have used both measures as proxies to quantify the performance of SEA campaigns \citep{YangGhose2010, Agarwal2011, AgarwalMukhopadhyay2016}. The third measure, 
\textit{cost-per-click}, is directly related to the relationship between the expenditure and outcomes, because it is an advertiser's expenditure on an ad divided by the number of clicks generated from the ad. For an advertiser, the actual CPC serves as a measure of how efficiently the advertising expenditure are generating clicks. 


Researchers have also investigated relationships among ad positions and consumer behavior measures \citep{GhoseYang2009}. For example, most studies agreed that CPC and CTR monotonically decrease when ad positions are lower \citep{GhoseYang2009, Agarwal2011, Jansen2013}. However, inconsistent findings exist for CVR. On one hand, \citet{GhoseYang2009} and \citet{Jansen2013} agreed that CVR is higher for ads at higher positions and decreases for lower ad positions. On the other hand, \citet{Agarwal2011} noticed that CVR could increase with lower ad positions. The reason is that, while ad positions do affect CTR, after a consumer clicks an ad, whether a conversion will occur depends mainly on the website and the product, instead of the ad position in the SERP \citep{Jansen2013}. 

Given the strong connection between ad positions and consumer behaviors, researchers have attempted to help advertisers place ads in the right positions. For example, \citet{GhoseYang2009} revealed that for search engines, bidding prices are more important than prior CTR for the final position of an advertisement. Keywords with more prominent positions are not necessarily the more profitable ones for advertisers. \citet{JeziorskiMoorthy2017} also studied the substitutional relationship between ad positions and advertisers' brand strength, and suggested that advertisers with strong brands do not necessarily need to bid for the highest position.




Due to the highly dynamic nature of SEA, many studies have also modelled dynamic advertising processes and related decisions, such as bidding for keywords. \citet{YaoMela2011} proposed a dynamic structural model to examine interactions among consumers, advertisers, and search engines, 
and revealed how ad positions affect advertisers' dynamic bidding behaviors. Advertisers' dynamic and strategic bidding behaviors in SEA could lead to a cyclical pattern on bidding prices: price-escalating phases are interrupted by price-collapsing phases. \citet{ZhangFeng2011} studied a cyclical bid adjustment model in a two-player setting, where an equilibrium bidding price and corresponding strategies for two advertisers can be obtained. 
\citet{AbhishekHosanagar2013} computed the optimal bids for keywords in an advertiser’s portfolio, while considering budget constraints and uncertainty in the decision environment. 
As online retailers adjust prices of their products and bidding prices on keywords of interest simultaneously and dynamically, \citet{Ye2014} studied how optimal bids and optimal prices change with inventory levels over time. 

In addition to bidding, another dynamic decision in SEA focuses on budget allocation over time. \cite{Yang2012} developed a framework with three levels of budgeting decisions through the life cycle of SEA campaigns, namely, allocation across search markets, temporal distribution over a series of time slots, and adjustment of the daily budget across keywords. \cite{Yang2015} also proposed an optimization-based budget allocation model for a monopolistic advertiser across two search markets over time. \citet{Zhang2014} studied the problem of daily budget adjustment over ad groups.


From the methodological perspective, our study is related to time-varying modeling \citep{Tan2012, Naik2015}. When dealing with longitudinal data, researchers often want to explicitly capture changes in the association between covariates and the outcome over time in a flexible manner. Thus \citet{Tan2012} introduced a time-varying coefficient (TVC) model--a special case of varying-coefficient model \citep{HastieTibshirani1993}. It has been used to explore the changing roles of regulatory regimes, marketing mailers, transaction characteristics and demographic factors on international trades and marketing outcomes \citep{Osinga2010, StremerschLemmens2009, Saboo2016}. 

Specifically, the TVC model has three characteristics that fit this study. \textit{First}, it is capable of estimating time-varying effects of covariates on the dependent variable. Thus TVC models are a generalized form of traditional linear regression models by incorporating time as the third dimension and representing coefficients of covariates with smoothly time-varying functions. 
\textit{Second}, compared to multi-level (or hierarchical) modeling (MLM) frameworks that can also capture temporal associations between time-varying covariates and the outcome, a TVC model is more flexible and could effectively reveal any arbitrary ``data-driven'' shapes of covariates' time-varying effects on the outcome, as long as coefficient functions are smooth (i.e., with no sudden jumps or break points)\footnote{A function is smooth if its first-order derivative function is continuous.}. In addition, in the TVC model framework, researchers can also specify a certain functional form when they have sufficient prior knowledge and evidence, while allowing others to change freely. By contrast, an MLM has to assume a specific form of coefficient functions (e.g., linear, quadratic, or cubic) for trajectory shapes. Admittedly,  
estimating a TVC model needs more data than a parametric model does \citep{Tan2012}. \textit{Third}, a TVC model can handle the co-existence of multiple covariates in the same model, including time-varying ones along with time-invariant ones. 


Overall, our research is distinct from the extant SEA research in the following ways: \textit{First}, this research focuses directly on sales generated from SEA campaigns and studies the ad-sales relationship. Compared to marketing outcome measures based on consumers' clicks, sales can more accurately and directly quantify advertisers' financial gains from SEA campaigns. \textit{Second}, we comprehensively investigate roles of a comprehensive group of factors, including the expenditure, carryover effects, consumer behaviors (e.g., CTR and CVR), and advertisement characteristics (e.g., ad position), in generating sales from SEA for the first time. More importantly, we study these factors' roles in a dynamic way so that we can reveal how their effects on sales change over time during an extended campaign period. \textit{Third}, we propose a time-varying response model for SEA and estimate its parameters using a large-scale dataset from a major E-Commerce retailer. Compared to advertising models in the literature, our model incorporates a quality-adjustment structure and is more appropriate for the dynamic context of SEA. 






\section{Data and Variables}
\subsection{Dataset}

This research uses a large-scale panel dataset collected from SEA campaigns by a large U.S. retailer, which offers a wide range of consumer electronics such as home appliances, air purifiers, etc. The retailer owns a large nationwide retail chain with brick-and-mortar stores and an electronic commerce website. The company has continuously conducted SEA campaigns over several years, and recorded data about SEA advertisements and online sales generated by these ads. The dataset we use is about SEA campaigns by this retailer during a 33-month period, spanning 4 calendar years. The dataset contains almost 7 million time-stamped records from nearly 40,000 key search phrases and almost 55,000 advertisements.

Each record in the dataset is about one advertisement on a given day. Specifically, a record includes keywords that triggered the ad, the number of impressions, the number of clicks, the average CPC, the number of conversions (i.e., purchase or orders), the total number of items ordered, and generated sales. Note that the search query of a keyword may lead to an impression (i.e., display) of a related ad, but not necessarily a click; a click may not lead to a conversion (i.e., an order), and an order may include one or more items.

We believe this dataset is appropriate for investigating the time-varying ad-sales relationship in SEA, because sales from SEA are available and the dataset covers a long time period that is sufficient to estimate a time-varying model \citep{Tan2012}. There are few empirical studies of SEA using a dataset that has such a large scale, covers such an long time span, or contains such a rich range of advertising and keyword attributes.

\subsection{Key Variables}
\label{subsect:keyvar}
Because this paper focuses on the ad-sales relationship, We directly use the number of products (in units) sold online ($Sales$) from each advertisement during a day as the dependent variable. There are several key independent variables whose effects on sales are of interests.

The first one is the expenditure ($AdExpenditure$) spent on an SEA advertisement on a given day. 
We also include three independent variables for consumer's click behaviors--\textit{CTR}, \textit{CVR}, and \textit{CPC}--and one independent variable for advertisement characteristics--ad position. At the first glance, it may seem that the transitive relationship from impressions, through clicks to conversions is simply linear in SEA. In other words, the number of clicks on an ad is the product of the number of impressions and CTR. Similarly, the number of conversions is the product of the number of clicks and CVR. However, such linear relationships do not necessarily hold, because CTR and CVR are not constants. Also, the relationship between the advertising expenditure and the number of clicks is essentially nonlinear because cost-per-click ($CPC$) also changes over time. 
Thus, we investigate the dynamic influence of each factor on sales over time.  

An important concept in SEA is the quality of an ad, which has a significant influence on the ad's performance \citep{KatonaZhu2018}. For search engines, an ad ranking mechanism that considers advertising quality facilitates better matching between advertisements and queries, and consequently improves revenue \citep{Feng2007, ChenStallaert2014}. For advertisers, a higher-quality ad means the advertiser can pay less for each click, so the same advertising budget can lead to more clicks and potentially higher sales. However, while search engines have widely adopted quality scores for ads, such scores are only available within search engine themselves. In addition, the exact formulas to calculate quality scores vary from one search engine to another, and remain trade secrets. 


Although advertisers have no access to their ads' quality scores and how such scores are calculated, Google Adwords reveals that quality scores mainly consider three factors\footnote{https://support.google.com/adwords/answer/6167118}: \textit{(1)} the expected CTR, which is also an independent variable in our model; \textit{(2)} advertising relevance, which indicates how closely an ad matches the intention of a consumer's search query or keyword(s); and \textit{(3)} the landing page experience. Note that the CVR does not affect an ad's quality score\footnote{https://www.ppchero.com/google-confirms-conversion-rates-have-no-effect-on-quality-score/}.

In addition, we also use four control variables related to search keywords: (1) $Brand$--whether the keywords contain any specific brand (e.g., ``\textbf{Apple} computer''); (2) $Retailer$--whether the keywords contain any specific retailer (e.g., ``\textbf{BestBuy} smartphone''); (3) $KLength$ (Length of keywords)--how many words are there in the search keywords (e.g., ``gift'' vs ``flower gift baskets'') \citep{GhoseYang2009}. Usually, it is more effective for an advertiser to choose brand-specific  retailer-specific, and longer (i.e., more specific) keywords \citep{RutzBucklin2011,AbhishekHosanagar2013}. (4) $Holiday$--whether the keywords contain any specific holiday. This is because advertisers often promote their products during holidays by using holiday keywords to raise consumers' desire to purchase (e.g., ``\textbf{Christmas} gift'').

Landing page experience refers to how relevant, transparent and easy-to-navigate the web page is for consumers who click an ad. In E-Commerce, a landing page generally corresponds to a product page, which shows details of the advertised product. A better landing page experience leads to a higher quality score\footnote{https://support.google.com/adwords/answer/2404196}. In our study, all ad-copies are for the same retailer, which means all the landing pages offer similar experience. Thus we treat landing page quality as a constant for all ads and skip it from our analysis.


See Table 1 for a list of all variables, along with their summary statistics, in this research. Table \ref{table:pairwise} illustrates the pairwise correlation among these variables.
 
\begin{table}[h!]\small
\caption{A Summary of Variables}
\label{table:variables}
\begin{center}
\begin{tabularx}{\textwidth}{X|X|l|l}
\hline
Variable& Description& Mean& Standard deviation \\
\hline
Sales($Sales$) & The total amount of sales (in units)& 16.271& 3521.487 \\ 
\hline
Lagged Sales($Sales_{t-1}$) & Sales from the previous time step (in units)& 16.271& 3521.487 \\ 
\hline
Advertising Expenditure ($AdExpenditure$) & Total spending (in dollars) on an ad during a day& 149.842&  2160.421\\
\hline
Ranking Position ($AdPosition$) & The ranking position of an ad on the SERP& 7.894&  10.922 \\
\hline
Cost-Per-Click ($CPC$) & The cost-per-click of an ad& 18.245&  53.637  \\
\hline
Click-Through-Rate ($CTR$) & The click-through-rate of an ad& 0.040&  0.131 \\
\hline
Conversion Rate ($CVR$) & The click-through-rate of an ad& 0.003& 0.045 \\
\hline
Length of Keywords ($KLength$) & The number of words in a keyword for an ad& 2.622&  0.803\\
\hline
Brand ($Brand$) & Binary variable--if associated keywords contain a specific brand& 0.175& 0.380 \\
\hline
Retailer ($Retailer$) &  Binary variable--if associated keywords contain a specific  retailer& 0.028&  0.167\\
\hline
Holiday ($Holiday$) &  Binary variable--if associated keywords contain a specific holiday& 0.003&  0.054\\
\hline
\end{tabularx}  
\end{center}
\end{table}

\begin{table}[h!]\small
\caption{Pairwise Correlation Coefficients Among Variables}
\label{table:pairwise}
\begin{center}
\includegraphics[scale=0.45] {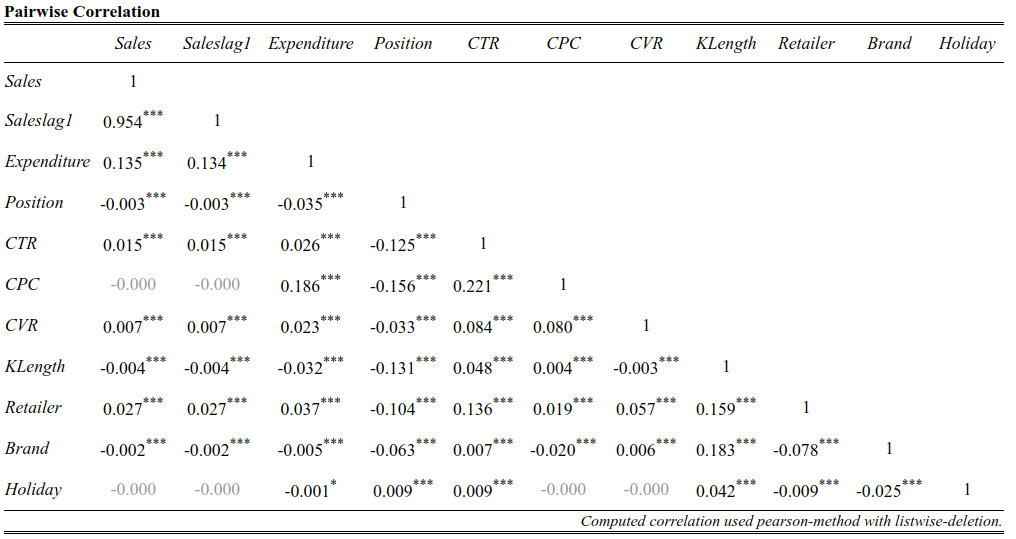}
\end{center}
\end{table}

\section{Model Development}
\subsection{Methodology} \label{tvc}
As defined in Equation \ref{eq:1}, our model has one dependent variable ($y_{ij}$) for advertising performance (i.e., sales from ads) along with a set of independent and control variables ($x_{ijk}$):
\begin{equation}
\label{eq:1}
y_{ij} = \beta_0(t_{ij} ) + \sum_{k=1}^{K}\beta_k(t_{ij})\cdot x_{ijk}  + \varepsilon _{ij} , \\ 
i=1,...,N; j=1,...,N_i ; k=1,...,K, 
\end{equation}
where $N$ represents the total number of subjects, $M_i$ is the total number of measurements for subject $i$ (i.e., an advertisement), and $K$ is the number of explanatory variables; $t_{ij} $ is the measurement time of the $j-th$ observation for the $i-th$ subject\footnote{Note that data can be unbalanced with different assessment time within and across individual subjects.}. 

$\beta_0(t_{ij})$ and $\beta_k(t_{ij})$ are the coefficient functions to be estimated: the intercept $\beta_0(t_{ij})$ represents the mean of $y$ when $x_k =0$ at time $t_{ij}$; The slope, $\beta_k(t_{ij})$, represents the strength and direction of the influence of $x_k$ on $y$ at time $t_{ij}$. Note that $\beta_0(t_{ij})$ and $\beta_k(t_{ij})$ are continuous coefficient functions of time $t$, such that their values change over time. 
Random errors $\varepsilon_{ij}$ in the above equation are assumed to be normally and independently distributed. Although time-varying parameters are treated as non-parametric functions, the class of TVC models is parametric for a specified time $t$. Thus the TVC model can be considered as conditionally parametric, representing a semi-parametric approach \citep{StremerschLemmens2009}. 

\subsection{A Time-varying SEA Response Model}\label{model}

\begin{figure}[h!]
\begin{displaymath}
\begin{array}{c}
\includegraphics[scale=0.6] {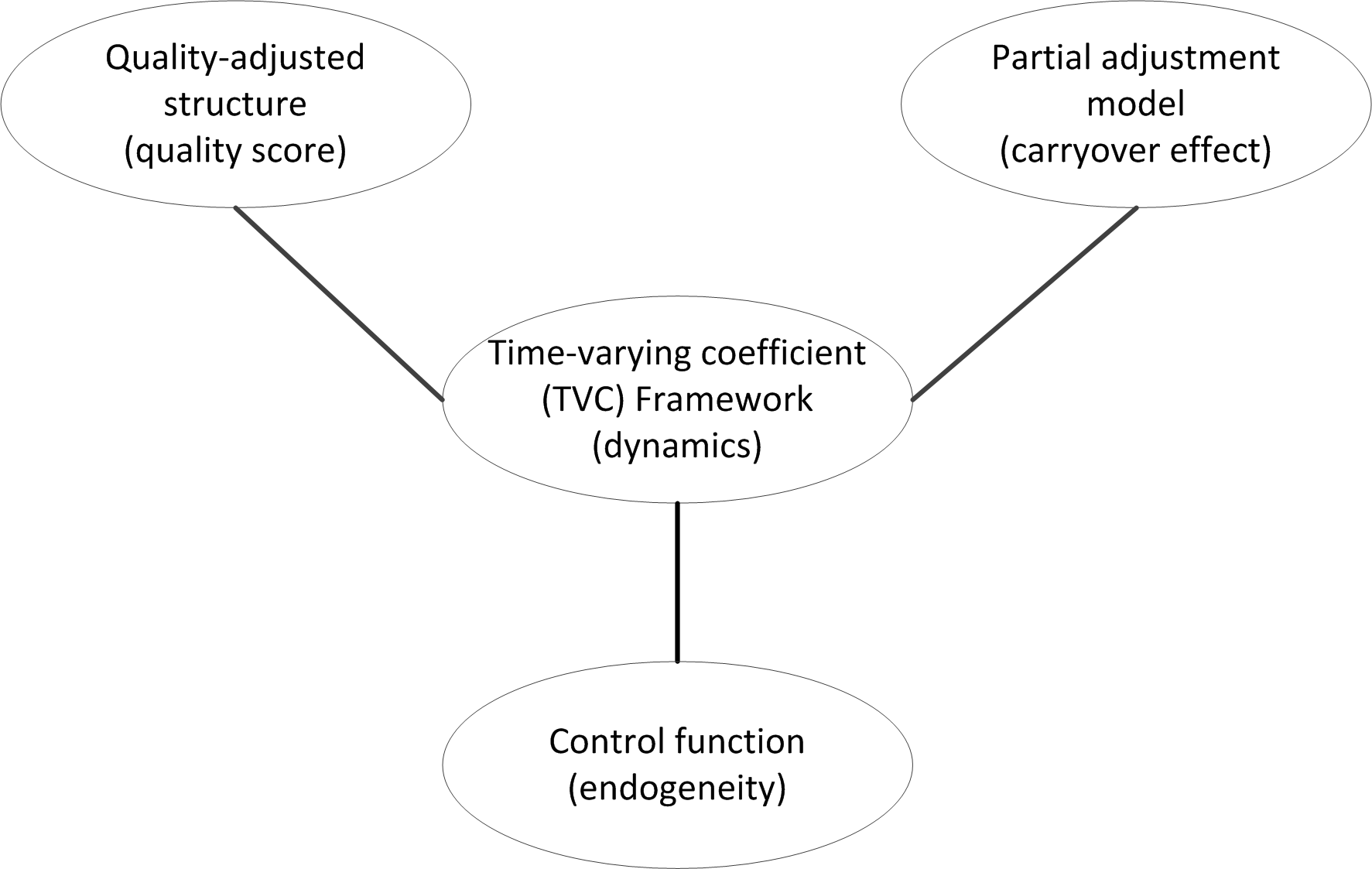}
\end{array}
\end{displaymath}
\caption[Caption for LOF]{The conceptual structure of the time-varying SEA response model\centering}
\label{fig:model}
\end{figure}

Given the nonlinear and temporal ad-sales relationship in SEA \citep{GhoseYang2009, Agarwal2011, Jansen2013}, 
we present an advertising response model for SEA markets in the TVC modeling framework (shown in Figure \ref{fig:model}), which does not assume any function forms for temporal trajectories of covariate coefficients in SEA.
We also incorporate a quality-adjusted structure \citep{Little1975, ParsonsSchultz1976} to account for the latent effect of advertising quality on ad performance 

\subsubsection{The Basic Model}
The basic model adopts the advertising response model by \citet{Arnold1987}. The model is an advertising spending function adjusted by a quality index based on the hedonic price theory (HPT) \citep{Ohta1975}. Because the outcome of a given amount of advertising spending depends on quality of the advertising copy, HPT can been naturally adopted to model the parsimonious process from advertising spending to market outcomes (e.g., sales). 
As discussed earlier in this paper, the quality of an ad explicitly affects the relationship between the advertising expenditure and sales, and needs to be controlled in our model. 
Therefore, following the Arnold model, we present a time-varying quality-adjusted response model for SEA as Equation \ref{eq:2}.

\begin{equation}
\label{eq:2}
Sales_{ij} =e^{\alpha_0(t_{ij})} \cdot (\psi _{ij} )^{\beta (t_{ij})} \cdot D_{ij} \cdot e^{\varepsilon_{ij}},  
\end{equation}
where $Sales_{ij}$ represents the number of products sold from advertisement $i$ at time $t_{ij}$; $\psi _{ij} $ is the advertising expenditure adjusted by the quality of advertisement $i$ measured at time $t_{ij}$. Note that advertising quality is latent and 
we will discuss how to estimate the quality-adjusted advertising spending function in the next subsection. $D_{ij}$ represents other covariates for sales. Details of them are in Subsection \ref{Determinants}. In addition, $\varepsilon_{ij}$ is the normally distributed error term at time $t_{ij}$; $\alpha_0(t_{ij)}$ and $\beta(t_{ij})$ will be estimated.

\subsubsection{The HPT-based Advertising Spending Function} \label{hpt}
Following the quality-adjusted market price in classical HPT \citep{Ohta1975}, we specify the quality-adjusted advertising expenditure function $\psi _{ij}$ from Equation \ref{eq:2} as below: 

\begin{equation}
\label{eq:3}
\psi _{t} =(B_{t} \prod_{k=1}^{K'}q_{kt}^{\tau_{k} } ),  
\end{equation}
where $B_{t} $ denotes the advertising spending (measured in dollar) at time $t$; $q_{kt} (k=1,..., K')$ is the value of advertisement attribute $k$ that determines an ad's quality score; $\prod_{k=1}^{K'} q_{kt}^{\tau_k}$ is thus the multi-dimensional quality index that is equivalent to the quality adjustment factor in \citep{ParsonsSchultz1976}(p.85), which adjusts the impact of the actual spending ($B_{t}$) based on the advertising quality.

Specifically in our case, as discussed in Subsection \ref{subsect:keyvar}, five attributes of an ad could affect its quality score: CTR, the length of keywords, and appearances of retailers, brands and holidays in keywords. 
Thus Equation \ref{eq:3} can be rewritten as a time-varying quality-adjusted advertising spending function defined in Equation \ref{eq:4}:

\begin{equation}
\label{eq:4}
\begin{array}{rl}
\psi_{ij} = & AdExpenditure_{ij} \cdot \theta_{ij}  \\ 
= &AdExpenditure_{ij} \cdot [{CTR_{ij} ^{\tau _{1} (t_{ij} )} }  \cdot \kappa _{ij}], \\ 
&with\ \kappa_{ij} = e^ {(KLength_i)^{\tau_2(t_{ij})} + (Retailer_i)^{\tau_3(t_{ij})} +(Brand_i)^{\tau_4(t_{ij})}  +  (Holiday_i)^{\tau_5(t_{ij})}}  ,
 \end{array}
\end{equation}
where $AdExpenditure_{ij}$ denotes the actual advertising expenditure
observed at time $t_{ij}$. $\theta_{ij} $ is the quality of advertisement
$i$ at time $t_{ij} $. It is a latent variable that is approximated by the product of time-dependent $CTR_{ij}$ (the CTR for advertisement $i$ measured at time $t_{ij}$) and time-invariant $\kappa_{ij}$, which represents the joint effects of four characteristics of keywords associated with advertisement $i$: $KLength_{i}$, $Retailer_{i}$, $Brand_{i}$ and $Holiday_{i}$.
Five coefficient functions $\tau_1(t_{ij})$, $\tau_2(t_{ij})$, $\tau_3(t_{ij})$, $\tau_4(t_{ij})$, and $\tau_5(t_{ij})$ will be estimated.

\subsubsection{Ad Position, CVR and CPC}\label{Determinants}

In addition to the expenditure, CTR, and keyword characteristics, three more independent variables--ad position, CVR, and CPC-- are included in $D_{ij}$, which is defined in Equation \ref{eq:D}:

\begin{equation}
\label{eq:D}
\begin{array}{rl}
D_{ij} = &\prod_{m=1}^{M} X_{m,ij}^{\lambda_m}  \cdot e^{\sigma_{ij} }  \\ 
= & e^{(AdPosition_{ij} )^{\lambda_1 (t_{ij} )}} \cdot (CPC_{ij} )^{\lambda_2 (t_{ij} )} \cdot {(CVR_{ij}) ^{\lambda_3 (t_{ij} )} } , \\
 \end{array}
\end{equation}
where $AdPosition_{ij} $ is the position of advertisement $i$ on SERPs, $CPC_{ij}$ and $CVR_{ij}$ are the cost-per-click and conversion rate of advertisement $i$, respectively, measured at time $t_{ij}$. $\lambda_1(t_{ij})$, $\lambda_2(t_{ij})$ and $\lambda_3(t_{ij})$ are the parameters to be estimated. 
 
\subsubsection{The SEA Response Model}

\noindent Substituting Equations (\ref{eq:4}) and (\ref{eq:D}) into Equation (\ref{eq:2}), we get

\begin{equation}
\label{eq:6}
\begin{array}{rl}
Sales_{ij} = &e^{\alpha_0 (t_{ij})} \cdot (AdExpenditure_{ij} \cdot [(CTR_{ij})^{\tau_1 (t_{ij})} \cdot e^{(KLength_i)^{\tau_2 (t_{ij})}} \cdot e^{(Retailer_i)^{\tau_3 (t_{ij})}} \\
& \cdot e^{(Brand_i)^{\tau_4 (t_{ij})}}  
\cdot e^{(Holiday_i)^{\tau_5 (t_{ij})}} ])^{\beta (t_{ij} )} \cdot e^{(AdPosition_{ij})^{\lambda_1 (t_{ij})}}   
\cdot (CPC_{ij})^{\lambda_2 (t_{ij})} \\
& \cdot {(CVR_{ij}) ^{\lambda_3 (t_{ij} )}  
\cdot e^{\varepsilon_{ij}}}
 \end{array}
\end{equation}

After taking natural logarithm transformations on numeric variables in Equation \ref{eq:6}, we obtain Equation \ref{eq:7}:

\begin{equation}
\label{eq:7}
\begin{array}{rl}
\ln Sales_{ij} = &\alpha_0 (t_{ij}) + \beta (t_{ij})(\ln AdExpenditure_{ij} + \tau_1 (t_{ij}) \ln CTR_{ij} + \tau_2 (t_{ij}) KLength_{i}\\ 
& + \tau_3 (t_{ij}) Retailer_i + \tau_4 (t_{ij})Brand_i + \tau_5 (t_{ij}) Holiday_{i})\\ 
& +\lambda_1 (t_{ij}) AdPosition_{ij} + \lambda_2 (t_{ij}) \ln CPC_{ij}  + \lambda_3 (t_{ij}) \ln CVR_{ij}
+\varepsilon_{ij}  ,
 \end{array}
\end{equation} 

\subsubsection{Adding Carryover Effects}

To account for the dynamic carryover effects of past advertising outcomes on current outcomes \citep{Tull1965, Clarke1976}, we add the time-lagged independent variable (i.e., $Sales_{ij-1}$) to Equation \ref{eq:7}, as in dynamic linear models. Specifically, we choose the partial adjustment model \citep{CaballeroEngel1992}. The model describes a dynamic response process where a variable adjusts over time to a series of desired values  \citep{Vanhonacker1983}. In other words, only some fraction of the desired adjustment is accomplished within a time period. In marketing research, the partial adjustment model has been widely adopted to describe the dynamic response process of sales to advertising and capture carryover effects of current advertising on future sales \citep{Clarke1973, ParsonsSchultz1976, Kohler2016}. As noted by \citet{Vanhonacker1983}, different from the carryover parameter in the Koyck model, the carryover parameter in the partial adjustment framework characterizes the complete dynamic nature of the advertising response. Then Equation \ref{eq:7} is transformed into Equation \ref{eq:8}:

\begin{equation}
\label{eq:8}
\begin{array}{rl}
\ln Sales_{ij} = &\eta (t_{ij}) \alpha_0 (t_{ij}) + (1 - \eta (t_{ij})) \ln Sales_{ij-1} + \eta (t_{ij})\beta (t_{ij})(\ln AdExpenditure_{ij} \\
&+\tau_1 (t_{ij}) \ln CTR_{ij} + \tau_2 (t_{ij}) KLength_i + \tau_3 (t_{ij} )Retailer_{ij} + \tau_4 (t_{ij}) Brand_{i} \\
&+\tau_5 (t_{ij})Holiday_{i}) + \eta (t_{ij})\lambda_1 (t_{ij} ) AdPosition_{ij} + \eta (t_{ij} )\lambda_2 (t_{ij} ) \ln CPC_{ij} \\
& + \eta (t_{ij} )\lambda_3 (t_{ij}) \ln CVR_{ij},
+\varepsilon_{ij} ,
 \end{array}
\end{equation}
where $\eta (t_{ij} )$ is the partial adjustment coefficient, and $(1-\eta (t_{ij} ))$ denotes the carryover effect at time $t_{ij} $. As $\eta (t_{ij} ) \to 1$, the effect of advertising on sales is mainly 
instantaneous and the carryover effect hardly exists; conversely, as $\eta (t_{ij} ) \to 0$, sales become increasingly persistent.

\subsubsection{Accounting for the Endogeneity of Budgeting Policies}
In general, advertisers need to allocate their expenditures over advertisements strategically to achieve marketing objectives (e.g., maximizing revenues from SEA campaigns) \citep{Yang2015}. Such budgeting policies could lead to the endogeneity problem \citep{Rossi2014}: the estimated effect of advertising budget on sales might be biased by the correlation between advertising budget and one or more unobserved latent factors in the error term of Equation (\ref{eq:8}). 
To account for such endogeneity, we use the control function approach, which has been widely used to eliminate the endogeneity bias with marketing mix variables in marketing research \citep{PetrinTrain2010, LuanSudhir2010}.

The control function approach is essentially a Two Stage Least Squares (2SLS) estimator. In the first stage, the correction term is estimated by regressing the advertising expenditure ($AdExpenditure_{ij}$) on a set of exogenous variables. In SEA, advertisers might plan their budget on an ad according to three factors \citep{Dinner2014}--search demand ($Demand_{ij}$), CTR, and CPC, as defined in Equation \ref{eq:9}. 

\begin{equation}
\label{eq:9}
\begin{array}{rl}
\ln AdExpenditure_{ij} &= \varphi _{ij}^{B} \cdot z_{ij}^{B} +\mu _{ij}^{B} \\
&= z_1 (t_{ij}) \ln Demand_{ij}  + z_2 (t_{ij})\ln CPC_{ij}  + z_3 (t_{ij}) \ln CTR_{ij} + \mu _{ij}^{B},  
\end{array}
\end{equation}
where 
$z_{ij}^{B}$ indicates the vector of exogenous variables (i.e., $Demand_{ij}$, $CTR_{ij}$, and $CPC_{ij}$) for 
the advertising expenditure, $\varphi_{ij}^{B} $ is the unknown parameter vector, and the random error $\mu _{ij}^{B}$ is assumed to be independently and normally distributed.

In the second stage, we include estimated residual
$\hat{\mu}_{ij}^{B}$ as an additional variable in Equation
\ref{eq:8}. We also remove $CPC_{ij}$ from the second stage model because CPC is related to sales indirectly via its relationship with the advertising expenditure.
This can be confirmed by the zero and insignificant correlation between CPC and sales, as compared to the positive and significant correlation between CPC and the advertising expenditure (See Table \ref{table:pairwise}).

Note that the advertising response model in Equation
\ref{eq:8} is a nonlinear regression with regard to parameters $\eta (t_{ij})$, $\alpha_0 (t_{ij})$, $\beta(t_{ij})$, $\tau_1 (t_{ij})$, $\tau_2 (t_{ij})$, $\tau_3 (t_{ij})$,$\tau_4 (t_{ij})$, $\tau_5 (t_{ij})$, $\lambda _1 (t_{ij} )$ and $\lambda _3 (t_{ij} )$. It is more convenient to consider it as a linear form for model estimation, and then use estimation results to identify these original parameters. The linear regression form of the final advertising response model is specified in Equation \ref{eq:10}. 
\begin{equation}
\label{eq:10}
\begin{array}{rl}
\ln SALES_{ij} = & \alpha_0^\ast (t_{ij}) + \gamma^\ast (t_{ij}) \ln Sales_{ij-1} + \beta^\ast (t_{ij}) \ln AdExpenditure_{ij} \\
& +\tau_1^\ast (t_{ij}) \ln CTR_{ij} + \tau_2^\ast (t_{ij}) KLength_i + \tau_3^\ast (t_{ij}) Retailer_i + \tau_4^\ast (t_{ij}) Brand_i \\
& + \tau_5^\ast (t_{ij}) Holiday_{i} + \lambda_1^\ast (t_{ij}) AdPosition_{ij} 
+ \lambda_3^\ast (t_{ij})  \ln  CVR_{ij}\\
& 
 +\alpha _1^\ast \mu _{ij}^{B} + \varepsilon_{ij}^\ast,\\
\end{array}
\end{equation}
where $\alpha_0^\ast (t_{ij}) = \eta (t_{ij}) \alpha_0 (t_{ij})$, 
$\gamma^\ast (t_{ij}) = 1 - \eta (t_{ij})$, 
$\beta^\ast (t_{ij}) = \eta (t_{ij} ) \beta (t_{ij} )$, 
$\tau_1^\ast (t_{ij}) = \eta (t_{ij} ) \beta (t_{ij} ) \tau_1 (t_{ij} )$, 
$\tau_2^\ast (t_{ij}) = \eta (t_{ij} ) \beta (t_{ij} ) \tau_2 (t_{ij} )$,
$\tau_3^\ast (t_{ij}) = \eta (t_{ij} ) \beta (t_{ij} ) \tau_3 (t_{ij} )$,
$\tau_4^\ast (t_{ij}) = \eta (t_{ij} ) \beta (t_{ij} ) \tau_4 (t_{ij} )$,
$\tau_5^\ast (t_{ij}) = \eta (t_{ij} ) \beta (t_{ij} ) \tau_5 (t_{ij} )$,
$\lambda_1^\ast (t_{ij}) =  \eta (t_{ij} )\lambda_1 (t_{ij} )$,
$\lambda_3^\ast (t_{ij}) = \eta (t_{ij} )\lambda_3 (t_{ij} )$. The budget correction term ($\mu _{ij}^{B}$) can be viewed as an additional explanatory variable in Equation \ref{eq:10}.

\section{Results}
Following previous studies \citep{StremerschLemmens2009, Tan2012}, we leverage the penalized spline (P-spline) smoothing approach introduced by \citet{EilersMarx1996} to estimate unknown coefficient functions in Equation (\ref{eq:10}). See Appendix \textbf{A2} for details on the P-spline approach. 

This section first presents results of endogeneity correction of advertising budget policies. Then we compare the fit of various model specifications, including the time-invariant model (baseline) and three variants of our proposed time-varying SEA response model. Finally, we report results of our advertising model and discuss potential implications. All covariates are standardized in our models.

\subsection{Budget Endogeneity Correction}

\begin{table}[h!]\small
\caption{First Stage Results of the Control Function for Budget Endogeneity Correction}
\label{table:controlFunc}
\begin{center}
\includegraphics[scale=0.6] {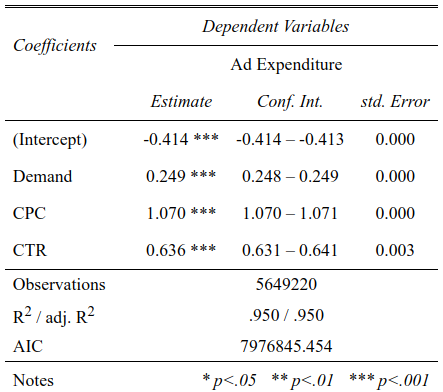}
\end{center}
\end{table}

Table \ref{table:controlFunc} presents the first stage results of our control function approach (i.e., Equation \ref{eq:9}), which corrects the potential endogeneity from strategical budget allocation policies. 
Results confirm advertisers' strategical budgeting decisions in SEA campaigns. Specifically, three exogenous variables--search demand, CPC and CTR--are all positive and statistically significant predictors for the advertising expenditure. 

Theoretically, in the PPC scheme, the influence of search demand, CPC and CTR on the advertising expenditure should be similar and close to $1.0$, because the expenditure can be computed as the product of these three factors. However, in our results, CPC appears to be the most influential factor for the advertising expenditure, followed by CTR and search demand. In other words, advertisers tend to emphasize on CPC and pay the least attention to search demand. 
This phenomenon is in line with the principles of information obtainability and least effort in information seeking behaviors \citep{JansenRieh2010}. On one hand, the principle of information obtainability states that, information that is more accessible to people is the more likely to be used by people, and vice versa. Similarly, according to the principle of least effort, when solving problems, a person tends to minimize her effort (over time). In the case of SEA, CPC has the highest obtainability for advertisers among the three factors, which is the most intuitive for them to understand and improve. 
This is because, for keywords with higher CPC (and bid prices), advertisers have to invest more in order to get sufficient opportunities to be displayed on SERPs and then clicked by search users.

By contrast, although advertisers often realize CTR's importance and have strong motivation to improve it, it is take much more time and effort to achieve a higher CTR and predict its temporal changes. In addition, precise information about search demands is challenging for advertisers to obtain during their campaigns. Even though some search engines or third-party companies (e.g., WordTracker) can provide potential search demand in a certain market, it is generally difficult for ordinary advertisers to predict the future search demand on daily basis and adjust advertising policies accordingly in a real-time way.
,

\subsection{Model Fit Comparisons}
Instead of a specific knot selection process, P-Spline-based approaches only need a large enough knot number (see Appendix for details), yet there is no agreement on the optimal number of knots. \citet{Wand2003} suggested the lower number between $35$ and $T/4$, where $T$ denotes the number of distinctive measurement times. \citet{Ruppert2002} recommended that $H$ around $10$ is enough to estimate monotonic functions and $H$ around $20$ is needed for complicated functions. Our dataset is unbalanced with different assessment time points within and across individual ads (i.e., $1 \leq T \leq 958$). In order to estimate parameters of our SEA response model (Equations \ref{eq:9} and \ref{eq:10}), we start with the B-Spline-based approach to fine-tune the analysis by incrementally increasing or decreasing the number of knots,   
and eventually use $H = 30$ in the P-Spline-based approach to estimate our model. We choose P-Splines over B-Splines because P-Splines can produce smoother estimates of the coefficient functions.

Next, we evaluate our time-varying model in terms of model fit by comparing it with several alternative specifications. The first alternative is a time-invariant model (MODEL-Time-Invariant), which treats coefficients of parameters in our full model (i.e., Equations \ref{eq:9} and \ref{eq:10}) as time-invariant constants. We also compare three variants of our time-varying model--Equations \ref{eq:9} and \ref{eq:10} specified with linear (MODEL-Time-Varying-linear), quadratic
(MODEL-Time-Varying-quadratic) and cubic (MODEL-Time-Varying-cubic) spline functions, respectively. Table \ref{table:modelfit} illustrates model fit statistics for these various specifications, including twice the negative of the residual log likelihood (-2 Res Log Likelihood), the Akaike information criterion (AIC) and the Bayesian information criterion (BIC).
 
\begin{table}[h!]\small
\caption{Model Fit Statistics of the Proposed Model and Alternative Specifications.}
\label{table:modelfit}
\begin{center}
\begin{tabular}{c|c|c|c|c}
\hline
 Model specifications & Trend specification & -2 Res Log Likelihood & AIC & BIC   \\
\hline
MODEL-Time-Invariant & NA & -4226454  & -4226431 & 4226268 \\
\hline
MODEL-Time-Varying-linear & linear spline & -4,290,931 & -4,290,335 & -4,286,298 \\
\hline
MODEL-Time-Varying-quadratic & quadratic spline & -4,297,012 & -4,296,394 & -4,296,394 \\
\hline
MODEL-Time-Varying-cubic & cubic spline & \textbf{-4,300,328} & \textbf{-4,299,688} & \textbf{-4,295,353} \\
\hline
\end{tabular}
\end{center}
\end{table}

As Table \ref{table:modelfit} shows, our time-varying advertising model specified with cubic splines provides the best fit, followed by MODEL-Time-Varying-quadratic and MODEL-Time-Varying-linear, while MODEL-Time-Invariant has the worst fit. In other words, including temporal dynamics helps time-varying models significantly improve their model fit compared to the time-invariant model. Theoretically, time-varying models break the study period into more fine-grained time intervals (rather than treat it as a single interval) and can reveal much more information about relationships between the explanatory variables and the dependent variable \citep{Tan2012}. Also, the best fit by cubic spline function, compared to the linear and quadratic functions, highlights the dynamic complexity of SEA markets \citep{Yang2018}.

\subsection{Parameter Estimates}
\subsubsection{Budget Endogeneity Correction}
\begin{table}[h!]\small
\caption{Estimated parameters for the time-invariant model with budget endogeneity correction.}
\label{table:ols}
\begin{center}
\includegraphics[scale=0.6] {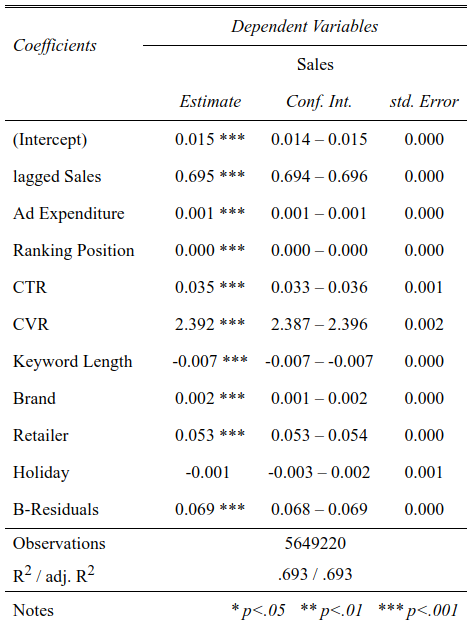}
\end{center}
\end{table}

To examine the effect of each covariate on sales, we also analyze estimated parameters from our time-invariant model with budget endogeneity correction (i.e., treating coefficients of Equations \ref{eq:9} and \ref{eq:10} as constants) because such coefficients are easier to understand and interpret than coefficient functions. Results in Tables \ref{table:ols} reveal two interesting findings: On one hand, the budget correction term ($\mu _{ij}^{B}$) has a statistically significant effect ($\alpha_1^\ast = 0.069$, $p < 0.001$), which justifies the addition of budget control function (i.e., Equation \ref{eq:9}) to the model. On the other hand, compared to other covariates, the budget correction term explains a substantial part of the variance in the dependent variable ($Sales$).
This suggests that, there are indeed some unobserved factors associated with advertisers' budgeting decisions. The positive effect of the budget correction term also implies that, without the budget correction process, parameter estimates of the advertising expenditure will be biased upwards, because the original model (in Equation \ref{eq:8}) omits unobserved factors that correlate with the advertising expenditure.

\begin{figure}[h!]
\begin{displaymath}
\begin{array}{c}
\includegraphics[scale=0.5] {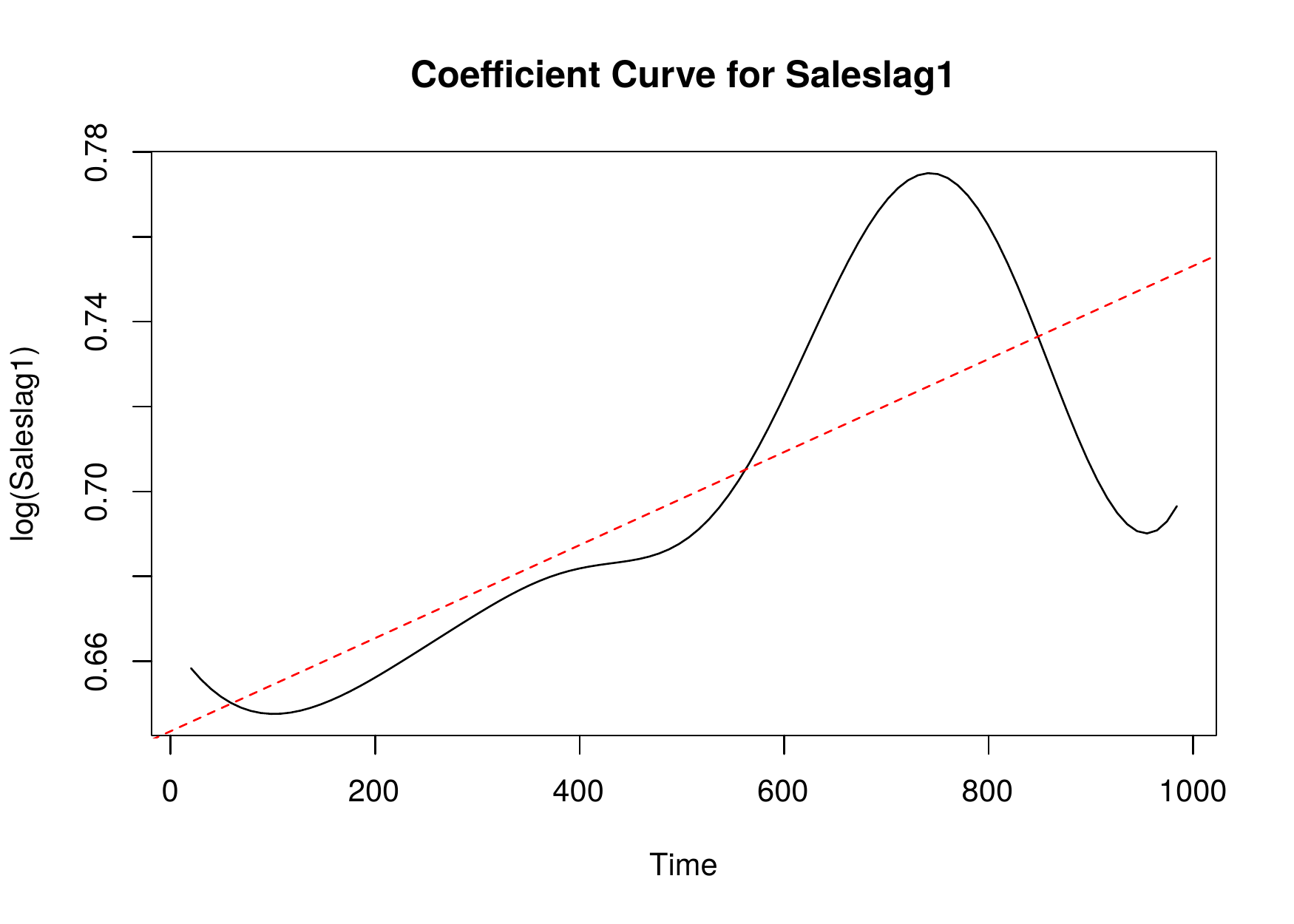}
\end{array}
\end{displaymath}
\caption[Caption for LOF]{The coefficient function of lagged sales\centering}
\label{fig:lagsales}
\end{figure}

\subsubsection{Carryover Effects}

From Table \ref{table:ols}, we can see that the variable for lagged sales ($Sales_{ij-1}$) has a statistically significant and positive effect on sales ($\gamma^\ast =0.695$, $p < 0.001$). 
Also, the coefficient of lagged sales fluctuates between $0.648$ and $0.774$ over time (Figure \ref{fig:lagsales}). 

Similar to traditional advertising, lagged sales remain a significant predictor for current sales in the context of SEA. This is in line with 
what has been reported in the literature (e.g., see \citet{Weiss1983, NaikRaman2003, Rosario2016}). 
The SEA carryover effect is also higher in SEA than those reported in traditional advertising channels (e.g., via newspapers, radio, TV, and billboards) \citep{Assmus1984,Arnold1987,Danaher2008}. Although \citet{Dinner2014} argued that the carryover effect is almost zero in SEA, their study was based on one apparel retailer, whose main revenue (nearly 85\%) is generated through offline channels. However, in the E-Commerce dataset we use, all sales driven by SEA are fulfilled online, which makes the estimation of carryover effects less biased.

From a temporal perspective, the carryover effect is persistent and strong in SEA over time. In general, carryover effect has an upward trend in SEA, as shown in Figure \ref{fig:lagsales}.
In addition, the trend is not monotonic for SEA, highlighting the complex dynamics in carryover effects.

\subsubsection{The Ad-sales Relationship}
\begin{figure}[h!]
\begin{displaymath}
\begin{array}{c}
\includegraphics[scale=0.5] {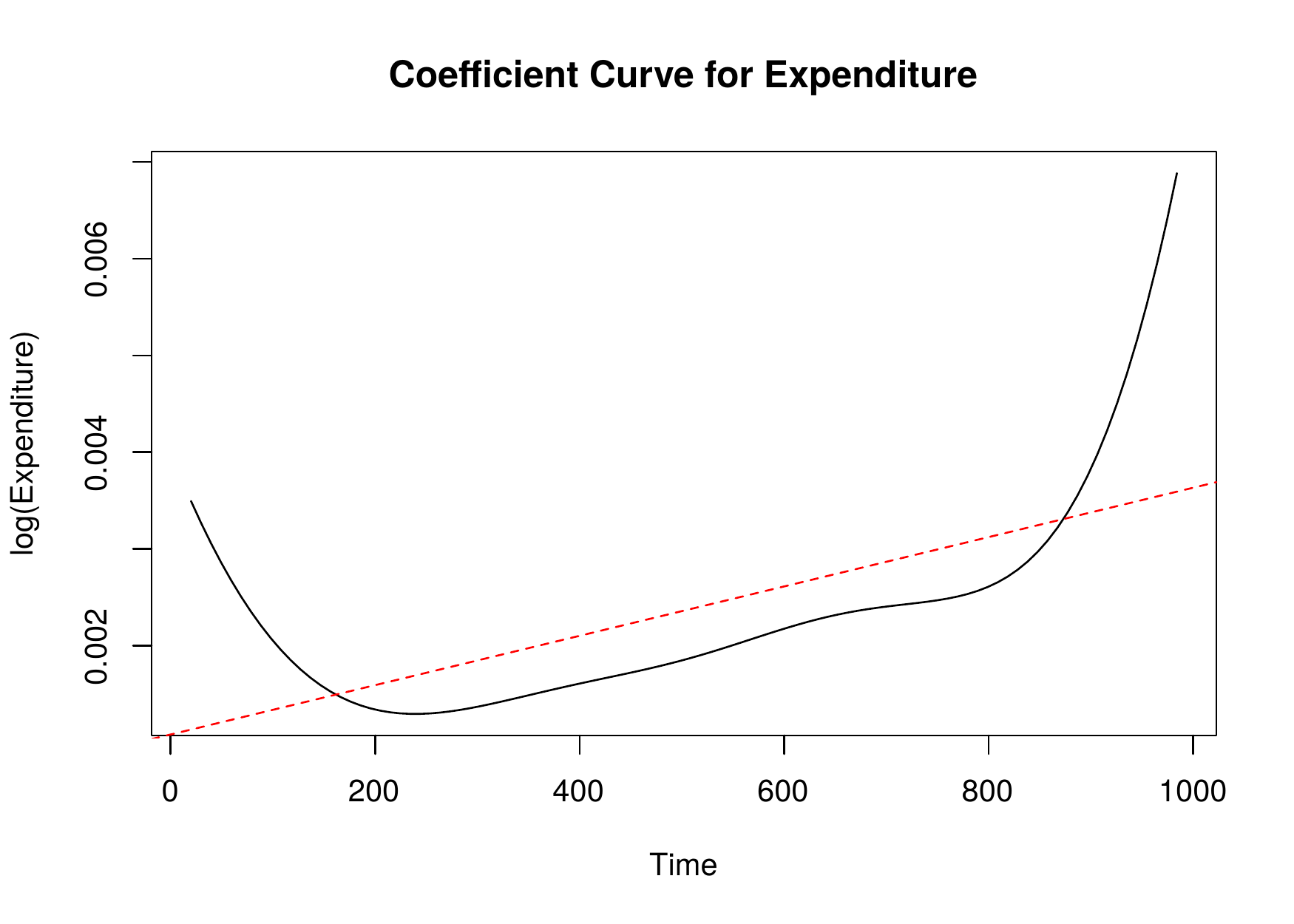}
\end{array}
\end{displaymath}
\caption[Caption for LOF]{The coefficient function of the advertising expenditure\centering}
\label{fig:expenditure}
\end{figure}

The coefficient of independent variable $AdExpenditure$ represents the short-term advertising elasticity. 
According to Table \ref{table:ols}, the advertising expenditure has a statistically significant and positive effect on sales ($\beta^\ast= 0.001$, $p < 0.001$). 
The coefficient of the advertising expenditure fluctuates between $0.001$ and $0.007$ over time (Figure \ref{fig:expenditure}) with a generally upward trend. 

Overall, despite the statistical significance, the magnitude of the current advertising expenditure's effect on sales is small. This contradicts the commonly-held views in  traditional advertising that the advertising expenditure is the major driving force to generate direct-response sales  \citep{Assmus1984,Sethuraman2011}. One possible reason for the difference is that millions of competing advertisers in SEA has led to more intense competitions \citep{Yang2016} and thus lower advertising elasticity \citep{Sethuraman2011, Dinner2014}. 

\subsubsection{Ad Positions, CTR and CVR}
\begin{figure}
\centering
\includegraphics[scale=0.45] {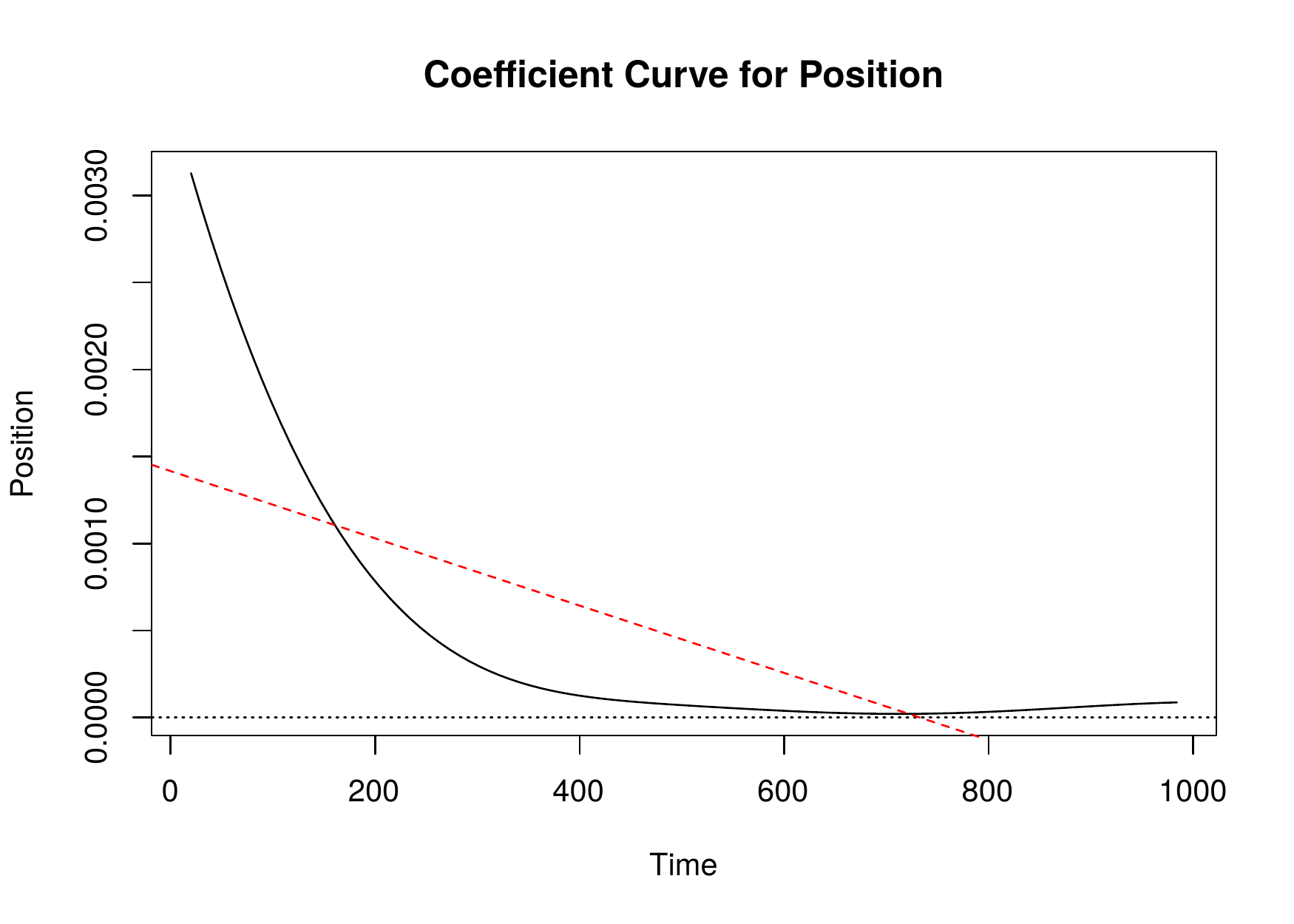}
\centering
\includegraphics[scale=0.45] {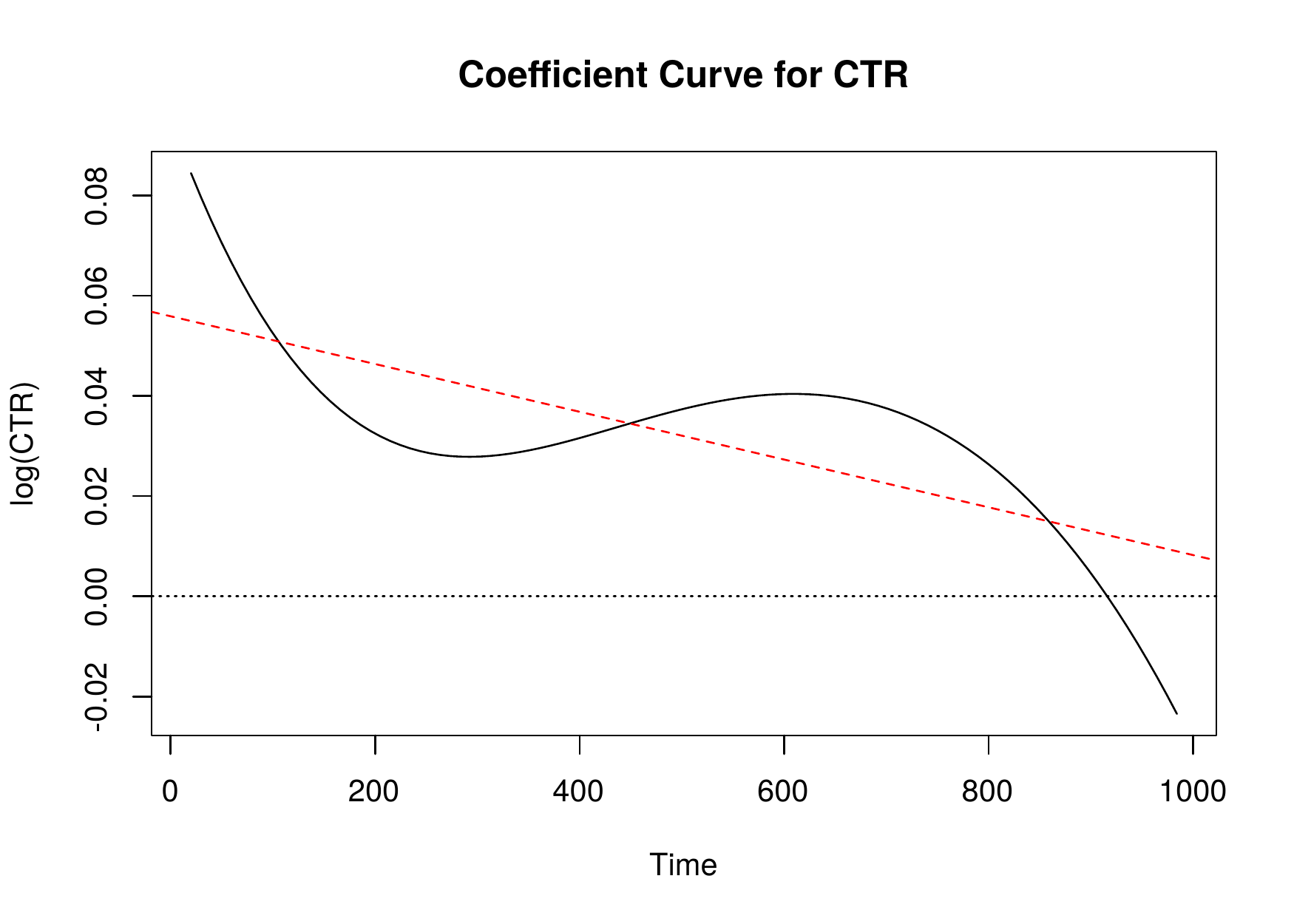}
\centering
\includegraphics[scale=0.45] {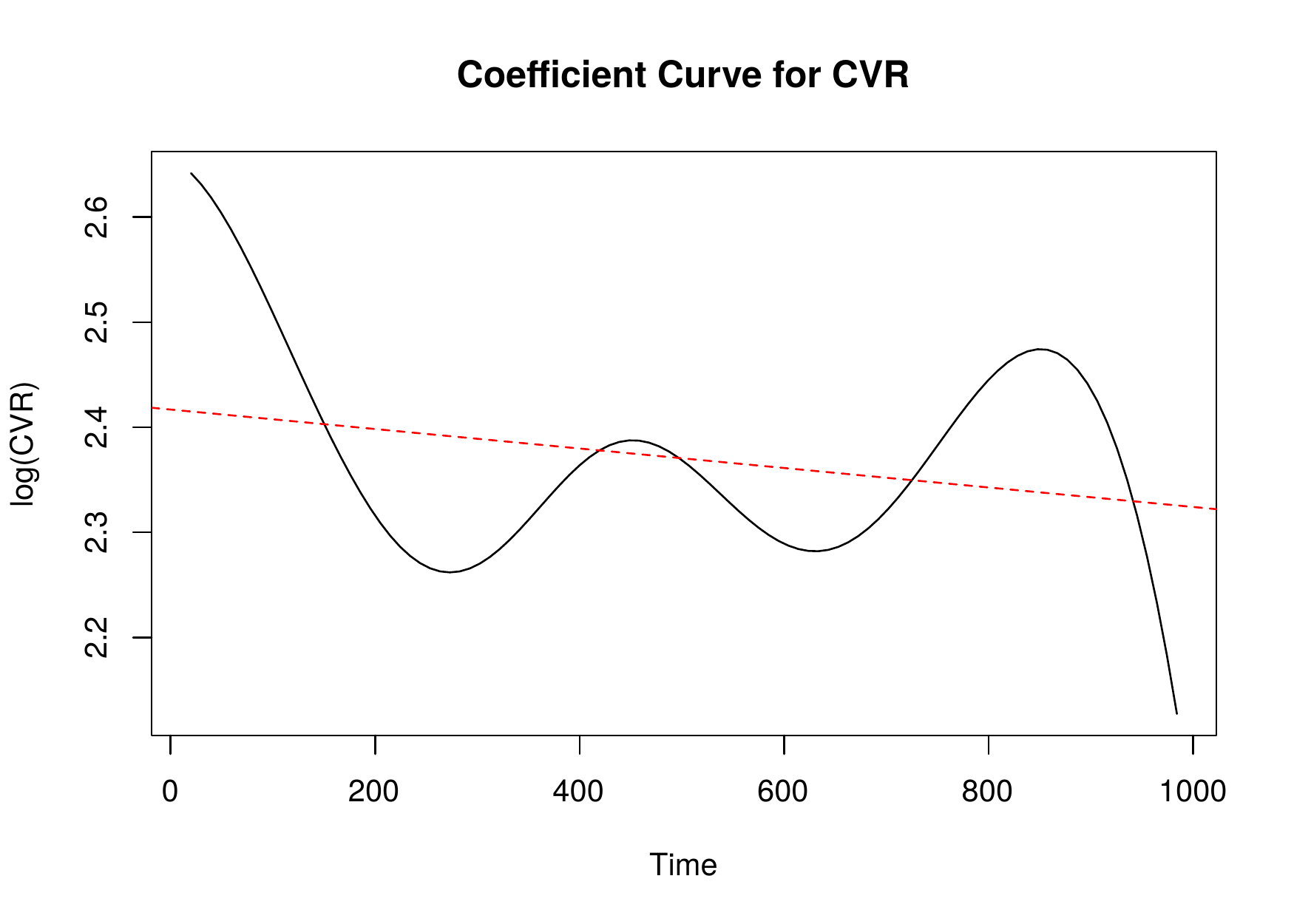}
\caption{Estimated coefficient functions of ad-specific factors}
\label{fig:adfactor}
\end{figure}

In this section, we investigate effects of advertisement characteristics (i.e., ad position) and consumer behavior measures (i.e., CTR and CVR) on sales (see Figure \ref{fig:adfactor}).

\textbf{Ad position}: Despite a weak pairwise correlation of $-0.003$ with sales (in Table \ref{table:pairwise}), ad position is turned out to be an insignificant predictor of sales ( $\lambda_1^\ast = 0.000$ and $p <0.001$) (see Table \ref{table:ols}) in our model. In other words, there is little difference in sales among ad positions on SERPs. One possible way to explain the insignificance is the theory of serial position effects. It states that it is easier for people to recall items at the beginning (primacy) or the end (recency) of a list of information than items in the middle
\citep{Ebbinghaus1913}.
Thus we also estimate a model with an additional quadratic term of ad position on ads listed on the first search engine results page (SERP). However, the new model still has coefficients of zeros for both ad position and its quadratic term. This again shows that an ad's position on SERPs does not affect the amount of sales it generates. This is a surprising finding that contradicts previous studies \citep{Agarwal2011,Jansen2013}. We will discuss the implications of this finding later in this paper.



\textbf{Click-through rate (CTR)} is statistically significant predictors of sales ($\tau_2^\ast = 0.035$, $p < 0.001$). 
Such a low coefficient is not surprising, because click-through is primarily a consequence of the brand building in online advertising. Moreover, Figure \ref {fig:adfactor} shows that the influence of CTR on sales declines over the promotion period. Its effect on sales even becomes negative at the final stage--as the campaign moves on, that is, ads with a lower CTR can produce more sales than those with a higher CTR. This is probably because at the later stage of a SEA campaign, certain search users get to know the advertiser and their ads better, and consider the advertiser's SEA ads as a quality source for products they desire. Consequently, when they do click an ad from the advertiser, they tend to purchase a higher amount of products \citep{Garcia-Molina2011}. 


\textbf{Conversion rate (CVR)} is statistically significant predictors of sales and their effect on sales ($\lambda_3^\ast = 2.392$, $p < 0.001$) is stronger than CTR. 
At the same time, CVR's influence on sales fluctuates over time. 

\subsubsection{Control Variables}
\begin{figure}
\centering
\includegraphics[scale=0.45] {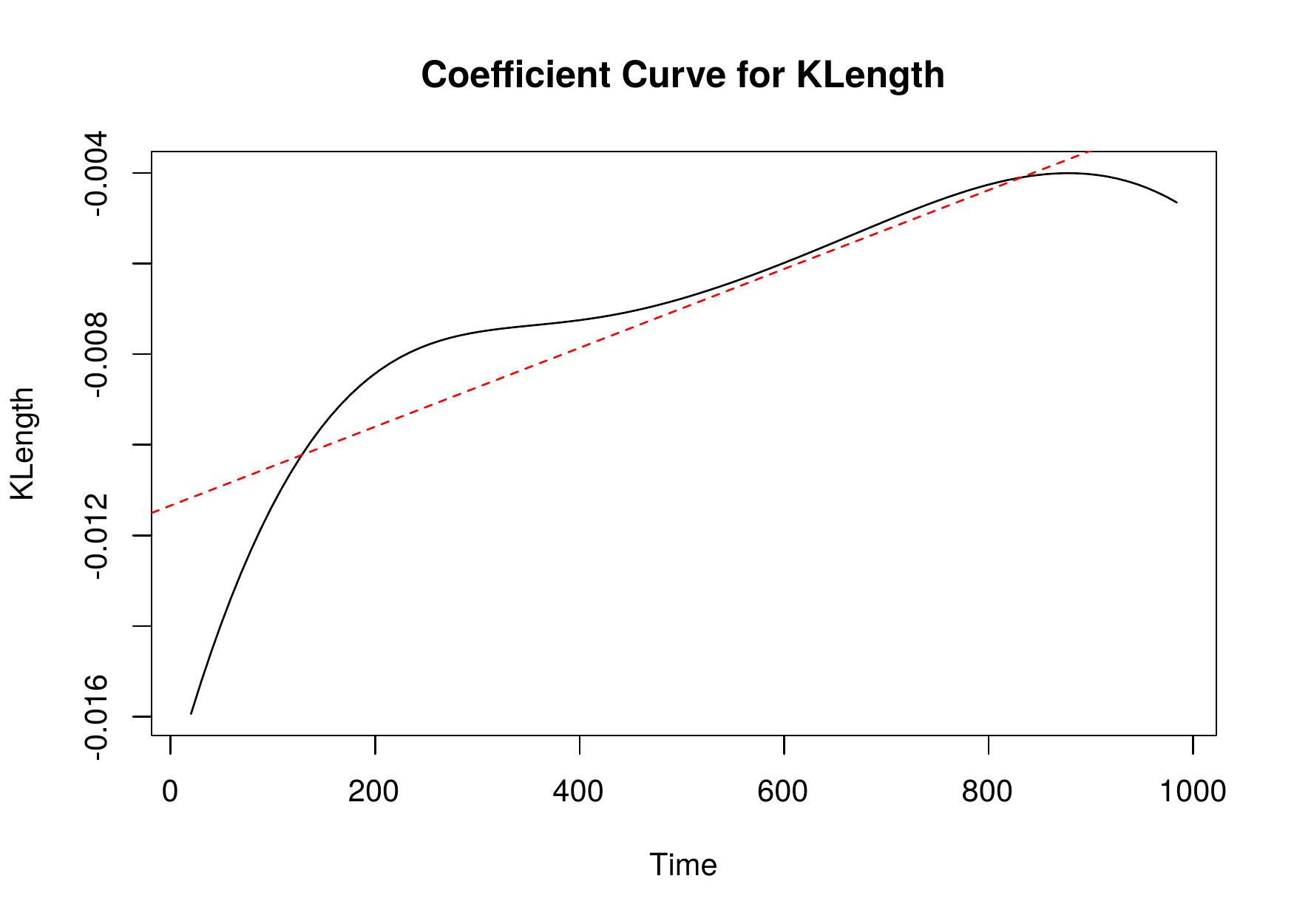}
\centering
\includegraphics[scale=0.45] {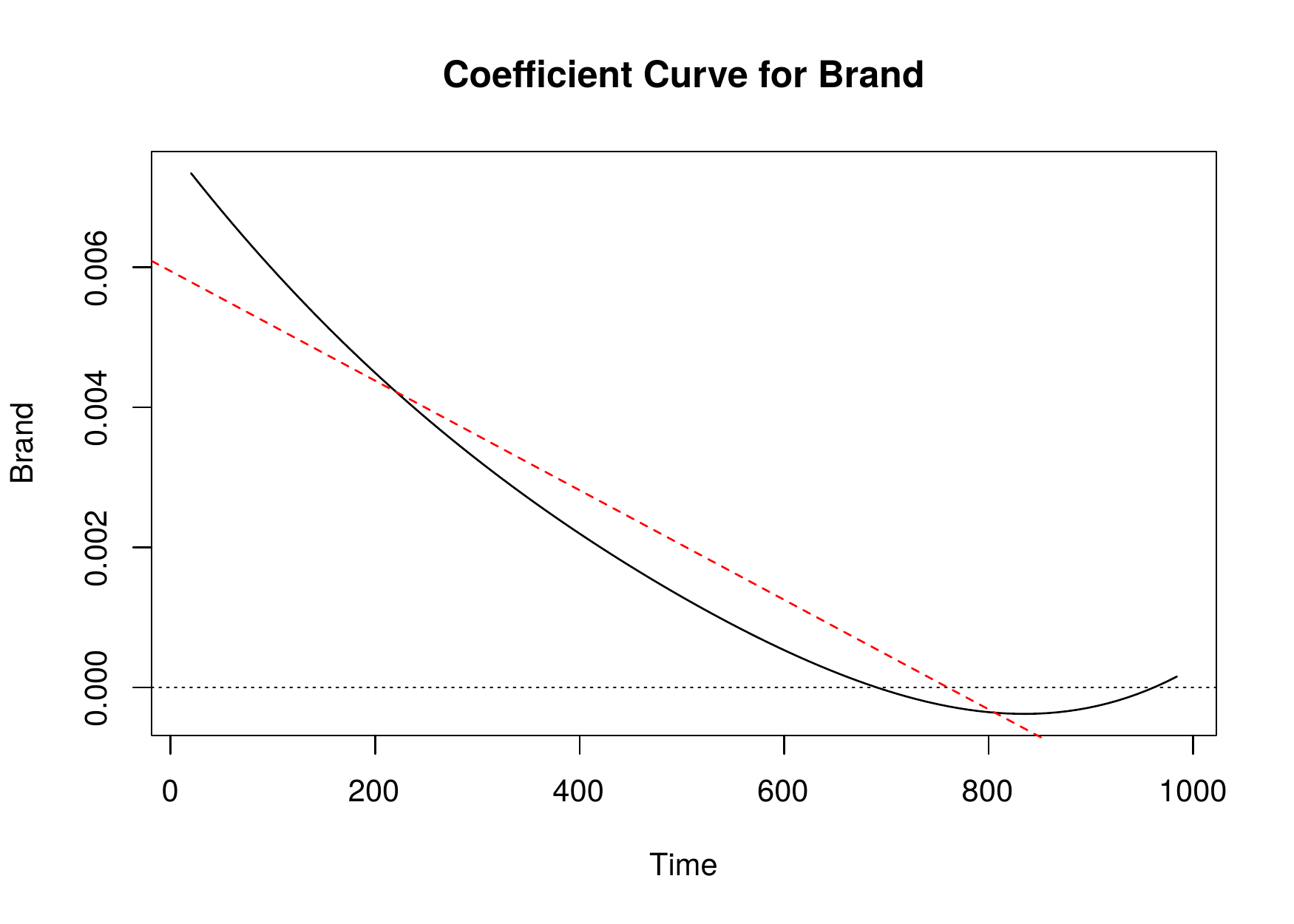}
\centering
\includegraphics[scale=0.45] {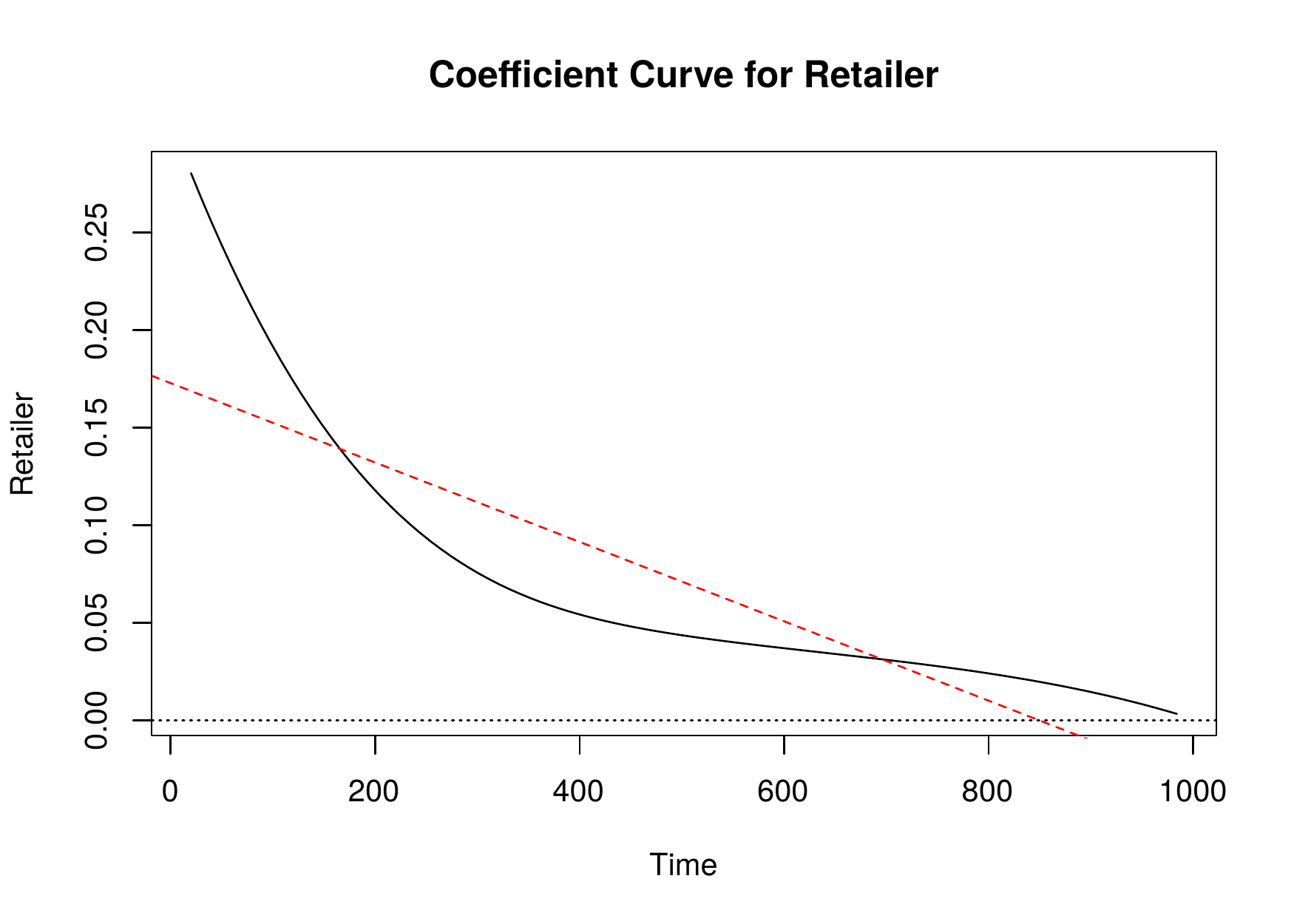}
\centering
\includegraphics[scale=0.45] {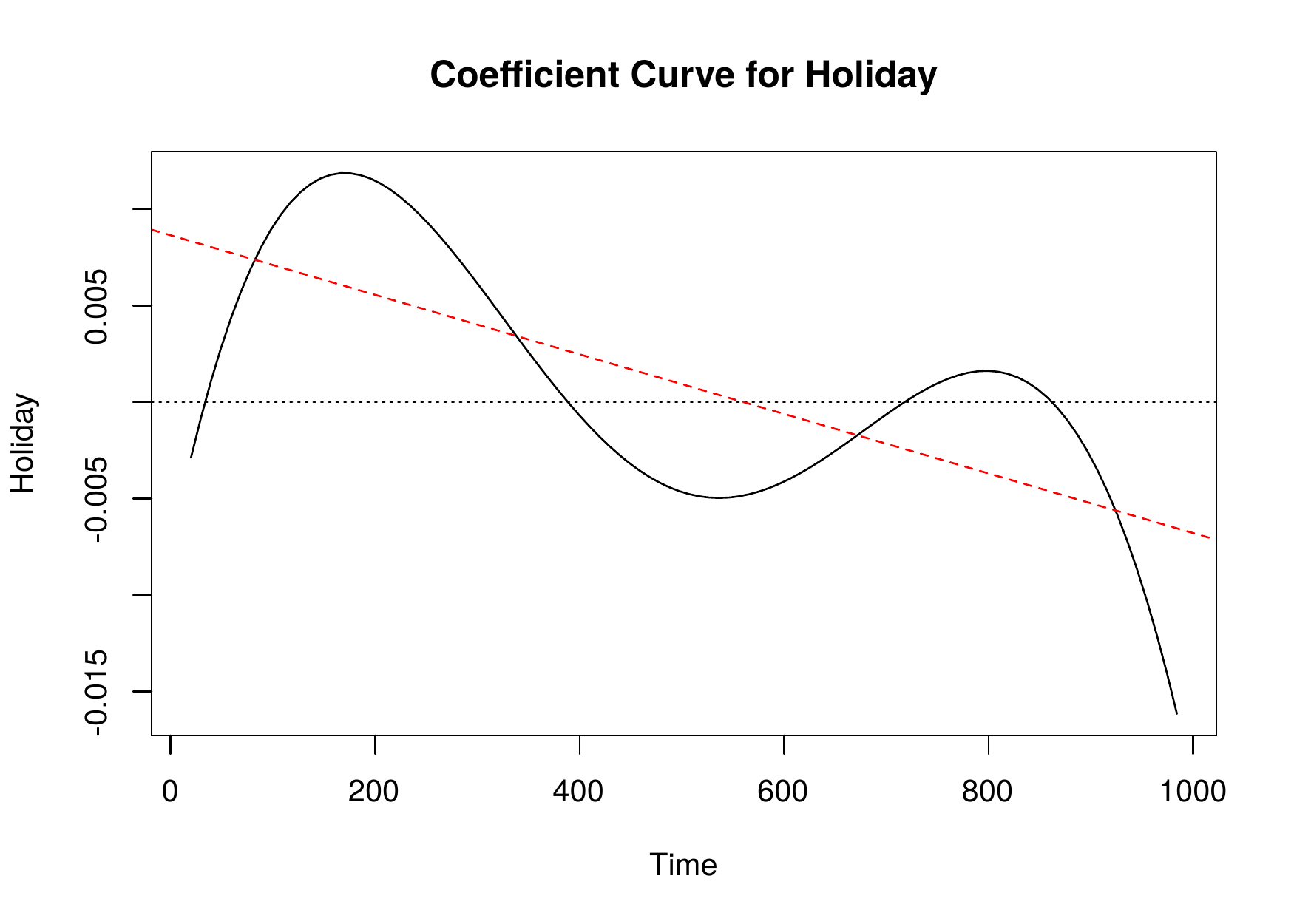}
\caption{Estimated coefficient functions of control variables}
\label{fig:keywordfactor}
\end{figure}

To represent latent advertising quality, we add control variables (Figure \ref{fig:keywordfactor}) to the model. 
\textbf{Keywords length} negatively influences sales ($\tau_2^\ast = -0.007$, $p < 0.001$)--a longer and more specific keyword leads to fewer sales than shorter and more general keywords. Even though consumers who search for more general keywords usually have lower CVR, a more general keyword can trigger much more clicks, leading to more sales.
As for what is in a keyword, containing a \textbf{brand} and containing a \textbf{retailer} in an ad are positive predictors of sales while containing a \textbf{holiday} is not. Also, all control variables' coefficient functions trend downward over time.

\section{Conclusions}

\subsection{Managerial Insights}
Our findings provide several managerial insights for SEA advertisers. First, they could help advertisers better understand the ad-sales relationships, especially the influence of various factors on sales, and how their influence changes over time. 
More importantly, our model entitles advertisers to predict advertising performance and allocate their advertising resources accordingly in a real-time fashion for their SEA campaigns\footnote{Our time-varying response model can be used to generate close-loop budget strategies over time via developing an optimal control model of budget planning \citep{Yang2015}. However, budget optimization is beyond the scope of this research.}
.

Second, for advertisers, focusing only on the direct and immediate effect of the expenditure on sales would underestimate the performance of their SEA campaigns in terms of sales, because SEA features a significant carryover effect that is more influential than the immediate effect. 
In other words, the effect of the advertising expenditure on sales would be overestimated without considering lagged sales (Table \ref{table:pairwise}). 
In practice, given the temporal dynamics in the carryover effect in SEA, advertisers could consider increasing/decreasing their advertising budget when the coefficient of the carryover effect is on the rise/decline, in order to get bigger ``bang of the buck''.

Third, our research also reveals the effects of advertisement characteristics and consumer behavior measures on SEA sales. Surprisingly, it is not practically meaningful for search advertisers to bid for higher ad positions on SERPs.
This challenges search engines' traditional stance that advertisers should always aim for higher ad positions on SERPs in order to maximize their advertising performance. Instead, our finding offers an alternative way for advertisers to gain higher return on investment (ROI)  by spending a little less and maintaining a lower ad position on SERPs.

Among measures of consumers behaviors, SEA advertisers should prioritize CVR over CTR, the latter is overstated in the current pay-per-click advertising scheme\footnote{https://www.en.advertisercommunity.com/t5/Performance-Optimization/Good-CTR-still-not-sales/td-p/1126984?nobounce\#}. Moreover, it is more important for an advertiser to improve its CVR during the initial stage of its SEA campaigns, because the influence of CVR gradually decline over time.

Last but not the least, advertisers can adjust their keyword selection strategies over time--they can focus more on shorter and more general keywords in the initial stage of a campaign, and then increase the portion of longer and more specific keywords over time. Meanwhile, it also helps to improve sales if an advertiser can include more retailer-specific and brand-specific keywords during the initial stage of a campaign.

\subsection{Theoretical Implications}
Besides practical implications for advertisers, our research also contributes to the literature of SEA. First, our research addresses an important gap in the literature of SEA by capturing the dynamic ad-sales relationship. In addition, our results reveal non-linearity in the temporal effects of various key factors in SEA on sales. This finding can inform future studies of temporal dynamics in SEA. 
Moreover, our model implicitly encapsulates the concept of advertising quality scores as a latent variable by adopting a quality-adjustment structure. This allows us to explore influence trajectories of advertising expenditure and various related factors on the expected market outcome (i.e., sales) over time.

Second, our research empirically compares advertising elasticity and carryover effect on SEA sales based on a large-scale dataset from a major U.S. E-Commerce retailer. Our results show that the carryover effect is stronger than elasticity and suggest that SEA is not a direct-response advertising medium. Instead, advertisers need to be patient and make a longer-term advertising investment before getting returns in sales. 
This also calls for more research on long-term strategies for SEA including budget allocation, bid pricing and keywords selections, because ordinary advertisers has little knowledge and time to operate such sophisticated and dynamical campaigns in the long run.

Third, our research also finds important patterns on how advertisers make budget decisions in SEA. We find that SEA advertisers mainly consider CPC, instead of CTR and search demand, when making such decisions. More explorations on advertisers' behaviors in different advertising schemes can help search engines improve their market design.

Last, our research helps to better understand inefficiency in the current SEA scheme. Prior studies \citep{GhoseYang2009, Agarwal2011} have found that higher positions on SERPs are not necessarily the more profitable ones for advertisers. Our research finds one potential reason for this. Advertising performance evaluation based on CTR is inevitability biased, because SEA campaigns experience a significant, positive and increasing carryover effect. In addition, the effects of CVR and CTR on sales make it possible to design a hybrid advertising scheme that combines pay-per-click and pay-per-action. 

\subsection{Limitations and Future Research}

We acknowledge several limitations of our research. First, similar to most, if not all, studies using advertising response models, we investigate the ad-sales relationship in SEA at the campaign level, rather than consumer behaviors at the individual level. The former is about how to allocate resources on advertising campaigns, while the latter focuses on how an advertiser should bid for a keyword in an auction against rivals. Our focus on the former means that our model cannot discern the heterogeneity inherent in behaviors of individual advertisers and their competitors.

Second, our study is limited by the dataset we used. For example, the dependent variable in our study is sales measured by the units of products sold from SEA campaigns. Sales is certainly important for advertisers and is often considered more valuable than clicks \citep{Sun2020}. However, return on investment that combines the advertising expenditure and profit from transactions can be a more straightforward way to measure an advertisers' financial gains. Besides monetary outcomes, some advertisers may also value the positive image of their brand gained from SEA campaigns. At the same time, even though our dataset is large in scale and covers an extended period of time, whether our conclusions apply to SEA in other contexts needs further investigations. 




This research can be extended in several ways. One interesting direction is to systematically understand the temporal pattern of each factor on sales, so that dynamic strategies for optimal resource allocation can be designed. 
We also plan to extend our model to include the continual bidding process and advertisers' behaviors at the individual level. In addition, time-varying interactions between advertisers and consumers should be an interesting topic to explore in
the field of SEA as well.


\clearpage


\clearpage

\section*{Appendix}

\subsection*{A1. Estimation of the Time-Varying Search Advertising Response Model} \label{estimate}

In the following we provide the estimation of the time-varying search
advertising response model described in Section \ref{model}. P-splines
have several very attractive merits. First, they do not impose any
assumption on the changing pattern of a given explanatory variable
with respect to time $t$ (e.g., linear, quadratic, or cubic), which
makes the estimated model immune to the misspecification problems
\citep{Tan2012}. Second, compared to smoothing approaches (e.g.,
regression splines, B-splines), P-splines have no boundary effects,
can conserve moments of data and have polynomial fits as limits, and
their computation are relatively inexpensive
\citep{EilersMarx1996}. Accordingly, P-splines have been widely used
in marketing literature on semi-parametric models (e.g.,
\citet{StremerschLemmens2009, Saboo2016}).

The general idea behind splines-based smoothers is that any smoothly
varying (coefficient) function (e.g., $f(t)$) defined on a certain
interval can be approximated by a linear combination of lower order
polynomial base functions. Specifically, the interval is partitioned
into $K+1$ smaller intervals, which are determined by $K$ dividing
points (i.e., knots), $\tau_1 ,\tau_2 ,..., \tau_K$; then we can
approximate $f(t)$ within each small interval $[\tau_{r}, \tau_{r+1}
), 0 \le r \le K$ with lower order polynomial functions. In the case
of time-varying coefficient functions \citep{Tan2012}, the $q$-order
truncated power basis can be specified as
 
\begin{equation} \label{eq:11} 
\begin{array}{l} 
t^0, t^1, t^2, ..., t^q, (t - \tau_1)_+^q, ..., (t - \tau_K)_+^q  \\ 
with{\rm \; }(t - \tau)_+^q = \left\{\begin{array}{l} {0{\rm \; if\; t}\le \tau } \\ {(t - \tau)^{q} {\rm \; if\; t}>\tau } \end{array}\right.  \end{array},        
\end{equation} 
where the first $q+1$ functions are the $0, 1, 2, ..., q$ order power
functions of $t$, and the other $K$ functions are truncated $q$ order
power functions determined by the $k$ knots, respectively.

In practice, researchers need to specify the number of knots (i.e.,
$K$). However, the number of knots is less crucial in P-splines based
estimation approaches, because they can optimally estimate the
coefficients using the linear mixed-effects model. Thus, theoretically
we can only choose a large enough $K$ (e.g., 10, which also depends on
the number of distinctive measurement times) for P-splines
\citep{Tan2012}.

As an example, the coefficient function $\alpha_0 (t_{ij})$ can be
approximately represented as
\begin{equation} \label{eq:12} 
\alpha_0 (t_{ij}) = a_0 +a_1 t_{ij} + a_2 t_{ij}^2 + \sum_{k=1}^{K} a_{q+k} (t_{ij} - \tau_k)_+^q.   
\end{equation} 

By substituting a set of coefficient functions of $t$ into the
original model to be estimated, we can get a linear regression model
with these basis functions (such as $1, t_{ij}, t_{ij}^2, ..., (t_{ij}
- \tau_k)_+^q$) as covariates and $a_0, a_1, a_2, ..., a_{q+k}$ as
coefficients, which can be easily estimated with ordinary least square
(OLS). P-splines combine B-splines with different penalties on
estimated coefficients, i.e., using ``a simple difference penalty on
the coefficients themselves of adjacent B-splines''
\citep{EilersMarx1996}, in order to address the overfitting
problem. The approach suggested by \citet{Ruppert2002} and
\citet{Wand2003} shrinks the coefficients of coefficient functions
(e.g., $a_{q+k}, k=1, 2, ..., K$ in Equation \ref{eq:10}) towards
zero, by minimizing the sum of SSE (sum of squared errors) and the
penalty term (defined as the summation of a series of products of
coefficients and corresponding tuning parameters), i.e., $SSE +
\lambda_1 \sum_{k=1}^{K} a_{q+k} (t_{ij} - \tau_k)_+^q + ...$. The
resulting optimal tuning terms (e.g., $\lambda_1$) balance the
tradeoff between the goodness of fit and smooth of the estimated
functions. Thus, the penalty term could prevent these coefficients
from being too large in absolute value. \citet{Wand2003} developed an
approach that treats these coefficients as random variable with normal
distribution, and expands the model to be estimated into a linear
mixed-effect model, which can be estimated with the restricted maximum
likelihood (REML) to the optimal balance. For more details on the
P-splines estimation of non- and semi-parametric models, see
\citep{EilersMarx1996, Ruppert2002, Wand2003}.


\begin{thebibliography}{00}\addtolength{\itemsep}{-1.5ex}
\bibitem[Abhishek and Hosanagar(2013)]{AbhishekHosanagar2013} Abhishek, Vibhanshu, and Kartik Hosanagar. (2013). ``Optimal bidding in multi-item multislot sponsored search auctions." \textit{Operations Research}, 61(4), 855-873.

\bibitem[Adaplo(2019)]{Adaplo2019}Adaplo. (2019). ``Google Shopping Campaigns Optimisation," Accessed September 3, 2019, \url{https://adaplo.com/google-shopping-optimisation}.

\bibitem[Anderson(2005)]{Anderson2005} Anderson Chris. (2005). ``Google's Long Tail," Stable URL (accessed May 10, 2017): \url{http://www.longtail.com/the_long_tail/2005/02/googles_long_ta.html}

\bibitem[Agarwal et al.(2011)]{Agarwal2011} Agarwal, Ashish, Kartik Hosanagar, and Michael D. Smith. (2011). ``Location, location, location: An analysis of profitability of position in online advertising markets," \textit{Journal of marketing research}, 48(6), 1057-1073.

\bibitem[Agarwal and Mukhopadhyay(2016)]{AgarwalMukhopadhyay2016} Agarwal, Ashish, and Tridas Mukhopadhyay. (2016). ``The Impact of Competing Ads on Click Performance in Sponsored Search." \textit{Information Systems Research}, 27(3), 538-557.

\bibitem[Archak et al.(2012)]{Archak2012} Archak, Nikolay, Vahab Mirrokni, and Shanmugavelayutham Muthukrishnan. (2012). ``Budget optimization for online campaigns with positive carryover effects." In International Workshop on Internet and Network Economics (pp. 86-99). Springer, Berlin, Heidelberg.

\bibitem[Arnold et al.(1987)]{Arnold1987} Arnold, Stephen J., Oum, Tae H., Pazderka Bohumir and Snetsinger Douglas W. (1987) ``Advertising quality in sales response models," \textit{Journal of Marketing Research}, 106-113.

\bibitem[Assmus et al.(1984)]{Assmus1984} Assmus, Gert, John U. Farley, and Donald R. Lehmann. (1984). ``How advertising affects sales: Meta-analysis of econometric results," \textit{Journal of Marketing Research}, 65-74.

\bibitem[Baadsgaard(2017)]{Baadsgaard2017}Baadsgaard, Jake. (2017). ``What Can CTR Tell Me About My Campaigns?" Accessed September 3, 2019, \url{https://www.disruptiveadvertising.com/adwords/what-is-ctr-click-through-rate/}

\bibitem[Rosario et al.(2016)]{Rosario2016}Babić Rosario, Ana, Francesca Sotgiu, Kristine De Valck, and Tammo HA Bijmolt. (2016). ``The effect of electronic word of mouth on sales: A meta-analytic review of platform, product, and metric factors," \textit{Journal of Marketing Research}, 53(3), 297-318.

\bibitem[Blake et al.(2015)]{Blake2015}Blake, Thomas, Chris Nosko, and Steven Tadelis. (2015). ``Consumer heterogeneity and paid search effectiveness: A large‐scale field experiment." Econometrica, 83(1), 155-174.

\bibitem[Caballero and Engel(1992)]{CaballeroEngel1992}Caballero, Ricardo J. and Eduardo MRA Engel. (1992). ``Beyond the partial-adjustment model," \textit{The American Economic Review}, 360-364.

\bibitem[Challis(2014)]{Challis2014}Challis, James. (2014). ``Short Term \& Long Term PPC Advertising Goals," Accessed September 3, 2019, \url{https://www.koozai.com/blog/pay-per-click-ppc/short-term-long-term-pay-per-click-advertising-goals/}

\bibitem[Chen and Stallaert(2014)]{ChenStallaert2014} Chen, Jianqing, and Jan Stallaert. (2014). ``An economic analysis of online advertising using behavioral targeting." \textit{Mis Quarterly}, 38 (2), 429-A7. 

\bibitem[Clarke(1973)]{Clarke1973} Clarke, Darral G. (1973). ``Sales-advertising cross-elasticities and advertising competition," \textit{Journal of Marketing Research}, 250-261.

\bibitem[Clarke(1976)]{Clarke1976} Clarke, Darral G. (1976). ``Econometric measurement of the duration of advertising effect on sales," \textit{Journal of Marketing Research}, 345-357.

\bibitem[Comscore(2008)]{Comscore2008}Comscore. (2008). ``Why Google's surprising paid click data are less surprising." Accessed September 3, 2019, \url{https://www.comscore.com/lat/Prensa-y-Eventos/Blog/Why-Google-s-surprising-paid-click-data-are-less-surprising}.

\bibitem[Da Silva(2018)]{DaSilva2018}Da Silva, C.N. (2018). ``Best Practices for Calculating and Mastering Your PPC Budget." Accessed September 3, 2019, \url{https://www.acquisio.com/blog/agency/best-practices-calculating-mastering-ppc-budget/}.


\bibitem[Danaher et al.(2008)]{Danaher2008} Danaher, Peter J., André Bonfrer, and Sanjay Dhar. (2008). ``The effect of competitive advertising interference on sales for packaged goods," \textit{Journal of Marketing Research}, 45(2), 211-225.


\bibitem[Dinner et al.(2014)]{Dinner2014} Dinner, Isaac M., Harald J. Heerde Van, and Scott A. Neslin (2014), ``Driving online and offline sales: The cross-channel effects of traditional, online display, and paid search advertising," \textit{Journal of Marketing Research}, 51(5), 527-545.

\bibitem[Du et al.(2015)]{Du2015}Du, Rex Yuxing, Ye Hu, and Sina Damangir. (2015). ``Leveraging trends in online searches for product features in market response modeling," \textit{Journal of Marketing}, 79(1), 29-43.

\bibitem[Ebbinghaus(1913)]{Ebbinghaus1913} Ebbinghaus, Hermann. (1913). \textit{Memory: A contribution to experimental psychology (No. 3)}. University Microfilms.

\bibitem[Eilers and Marx(1996)]{EilersMarx1996} Eilers, Paul HC, and Brian D. Marx. (1996). ``Flexible smoothing with B-splines and penalties," \textit{Statistical science}, 89-102.

\bibitem[Feng et al.(2007)]{Feng2007} Feng, Juan, Hemant K. Bhargava, and David M. Pennock. (2007). ``Implementing Sponsored Search in Web Search Engines: Computational Evaluation of Alternative Mechanisms." \textit{INFORMS Journal on Computing}, 19(1), 137-148.

\bibitem[Fischer et al.(2011)]{Fischer2011} Fischer, Marc,  Albers, Sönke, Wagner, Nils and Frie, Monika. (2011). ``Dynamic Marketing Budget Allocation Across Countries, Products, and Marketing Activities," \textit{Marketing Science}, 30(4): 568-585.

\bibitem[Garcia-Molina et al.(2011)]{Garcia-Molina2011} Garcia-Molina, Hector, Georgia Koutrika, and Aditya Parameswaran. (2011). ``Information seeking: convergence of search, recommendations, and advertising," \textit{Communications of the ACM}, 54(11), 121-130.

\bibitem[George(2019)]{George2019}George, Trevor. (2019). ``How to use Amazon advertising’s dynamic bidding feature." Accessed September 3, 2019, \url{https://searchengineland.com/how-to-use-amazon-advertisings-dynamic-bidding-feature-320505}. 

\bibitem[Ghose and Yang(2009)]{GhoseYang2009} Ghose, Anindya, and Sha Yang. (2009). ``An empirical analysis of search engine advertising: Sponsored search in electronic markets," \textit{Management Science}, 55(10), 1605-1622. 


\bibitem[GoogleAdwords(2019)]{GoogleAdwords2019}Google Adwords. (2019). ``About Quality Score." Accessed September 3, 2019, \url{https://support.google.com/google-ads/answer/7050591?hl=en}.

\bibitem[Hanssens et al.(2003)]{Hanssens2003}Hanssens, Dominique M., Leonard J. Parsons, and Randall L. Schultz. (2003). \textit{Market response models: Econometric and time series analysis (Vol. 12)}. Springer Science and Business Media.

\bibitem[Hastie and Tibshirani(1993)]{HastieTibshirani1993} Hastie, Trevor, and Robert Tibshirani. (1993). ``Varying-coefficient models," \textit{Journal of the Royal Statistical Society. Series B (Methodological)}, 757-796.


\bibitem[Jansen and Rieh(2010)]{JansenRieh2010}Jansen, Bernard J., and Soo Young Rieh. (2010). ``The seventeen theoretical constructs of information searching and information retrieval," \textit{Journal of the Association for Information Science and Technology}, 61(8), 1517-1534.


\bibitem[Jansen et al.(2013)]{Jansen2013} Jansen, Bernard J., Zhe Liu, and Zach Simon. (2013). ``The effect of ad rank on the performance of keyword advertising campaigns," \textit{Journal of the american society for Information science and technology}, 64(10), 2115-2132.

\bibitem[Jeziorski and Moorthy(2017)]{JeziorskiMoorthy2017}Jeziorski, P., and Moorthy, S. (2017). Advertiser prominence effects in search advertising. Management Science, 64(3), 1365-1383.

\bibitem[Johnson et al.(2017)]{Johnson2017}Johnson, Garrett, Randall A. Lewis, and Elmar Nubbemeyer. (2017). ``The online display ad effectiveness funnel \& carryover: Lessons from 432 field experiments." Available at SSRN 2701578.

\bibitem[Katona and Zhu(2018)]{KatonaZhu2018} Katona, Zsolt, and Yi Zhu (2018), ``Quality score that makes you invest," (Feb 10, 2018). Available at SSRN. \\ \url{https://ssrn.com/abstract=2954707}

\bibitem[Köhler et al.(2016)]{Kohler2016}Köhler, Christine, Murali K. Mantrala, Sönke Albers, and Vamsi K. Kanuri. (2016). ``A meta-analysis of marketing communication carryover effects," \textit{Journal of Marketing Research}, 54 (6), 990-1008.

\bibitem[Little(1975)]{Little1975} Little, John DC. (1975). ``BRANDAID: A marketing-mix model, part 1: Structure," \textit{Operations Research}, 23(4), 628-655. 

\bibitem[Luan and Sudhir(2010)]{LuanSudhir2010} Luan, Y. Jackie, and K. Sudhir. (2010). ``Forecasting marketing-mix responsiveness for new products," \textit{Journal of Marketing Research}, 47(3), 444-457.


\bibitem[Membrillo(2018)]{Membrillo2018}Membrillo, Alex. (2018). ``13 PPC Trends to Increase Your CTR and Sales." Accessed September 3, 2019, \url{https://www.cardinaldigitalmarketing.com/blog/13-ppc-trends-to-increase-your-ctr-and-sales/}

\bibitem[Naik and Raman(2003)]{NaikRaman2003}Naik, Prasad A., and Kalyan Raman. (2003). ``Understanding the impact of synergy in multimedia communications," \textit{Journal of Marketing Research}, 40(4), 375-388.

\bibitem[Naik(2015)]{Naik2015} Naik, Prasad A. (2015). ``Marketing dynamics: A primer on estimation and control," \textit{Foundations and Trends® in Marketing}, 9(3), 175-266.

\bibitem[Ohta(1975)]{Ohta1975} Ohta, Makoto. (1975). ``Production technologies of the US boiler and turbogenerator industries and hedonic price indexes for their products: a cost-function approach," \textit{Journal of Political Economy}, 83(1), 1-26.

\bibitem[Osinga et al.(2010)]{Osinga2010} Osinga, Ernst C., Peter SH Leeflang, and Jaap E. Wieringa. (2010). ``Early marketing matters: a time-varying parameter approach to persistence modeling," \textit{Journal of Marketing Research}, 47(1), 173-185.

\bibitem[Pabich(2011)]{Pabich2011}Pabich, Nathan, (2011). ``5 Long Term Benefits of PPC Advertising," Accessed September 3, 2019, \url{https://www.digitalthirdcoast.com/blog/5-long-term-benefits-of-ppc-advertising}.


\bibitem[Parsons and Schultz(1976)]{ParsonsSchultz1976} Parsons, Leonard J., and Randall L. Schultz. (1976). \textit{Marketing models and econometric research}. New York: North-Holland Publishing Company.

\bibitem[Petrin and Train(2010)]{PetrinTrain2010} Petrin, Amil, and Kenneth Train. (2010). ``A control function approach to endogeneity in consumer choice models," \textit{Journal of marketing research}, 47(1), 3-13.

\bibitem[Rossi(2014)]{Rossi2014}  Rossi, Peter E. (2014). ``Even the rich can make themselves poor: A critical examination of IV methods in marketing applications," \textit{Marketing Science}, 33(5), 655-672.

\bibitem[Ruppert(2002)]{Ruppert2002} Ruppert, David. (2002). ``Selecting the number of knots for penalized splines," \textit{Journal of computational and graphical statistics},11(4), 735-757.

\bibitem[Rutz and Bucklin(2010)]{RutzBucklin2011} Rutz, Oliver J., and Randolph E. Bucklin. (2011). ``From generic to branded: A model of spillover in paid search advertising," \textit{Journal of Marketing Research}, 48(1), 87-102.

\bibitem[Saboo et al.(2016)]{Saboo2016} Saboo, Alok R., V. Kumar, and Insu Park. (2016). ``Using Big Data to Model Time-Varying Effects for Marketing Resource (Re) Allocation," \textit{MIS Quarterly}, 40(4), 911-939.


\bibitem[Sethuraman et al.(2011)]{Sethuraman2011}Sethuraman, Raj, Gerard J. Tellis, and Richard A. Briesch. (2011). ``How well does advertising work? Generalizations from meta-analysis of brand advertising elasticities," \textit{Journal of Marketing Research}, 48(3), 457-471.

\bibitem[Stremersch and Lemmens(2009)]{StremerschLemmens2009} Stremersch, Stefan, and Aurélie Lemmens. (2009). ``Sales growth of new pharmaceuticals across the globe: The role of regulatory regimes," \textit{Marketing Science}, 28(4), 690-708.

\bibitem[Sun et al.(2020)]{Sun2020} Sun, Haoyan, Ming Fan, and Yong Tan. (2020). ``An Empirical Analysis of Seller Advertising Strategies in an Online Marketplace." \textit{Information Systems Research}, 31 (1), 37-56.

\bibitem[Tan et al.(2012)]{Tan2012} Tan, Xianming, Shiyko, Mariya P., Li, Runze, Li, Yuelin and Dierker, Lisa. (2012). ``A time-varying effect model for intensive longitudinal data," \textit{Psychological methods}, 17(1), 61-77.

\bibitem[Tull(1965)]{Tull1965} Tull, Donald S. (1965). ``The carry-over effect of advertising," \textit{Journal of Marketing}, 46-53.


\bibitem[Vanhonacker(1983)]{Vanhonacker1983} Vanhonacker, Wilfried R. (1983). ``Carryover effects and temporal aggregation in a partial adjustment model framework," \textit{Marketing Science}, 2(3), 297-317.


\bibitem[Wand(2003)]{Wand2003} Wand, Matt P. (2003). ``Smoothing and mixed models," \textit{Computational statistics}, 18(2), 223-249.

\bibitem[Weiss et al.(1983)]{Weiss1983}Weiss, Doyle L., Charles B. Weinberg, and Pierre M. Windal. (1983). ``The effects of serial correlation and data aggregation on advertising measurement," \textit{Journal of Marketing Research}, 268-279.

\bibitem[Yang and Ghose(2010)]{YangGhose2010} Yang, Sha and Ghose, Anindya. (2010). ``Analyzing the relationship between organic and sponsored search advertising: Positive, negative, or zero interdependence?" \textit{Marketing Science}, 29(4), 602-623. 

\bibitem[Yang et al.(2012)]{Yang2012} Yang, Yanwu, Jie Zhang, Rui Qin, Juanjuan Li, Fei-Yue Wang, and Wei Qi. (2012). ``A Budget optimization framework for search advertisements across markets," \textit{IEEE Transactions on Systems, Man, and Cybernetics. Part A: Systems and Humans}, 42(5), 1141-1151. 

\bibitem[Yang et al.(2015)]{Yang2015} Yang, Yanwu, Zeng, Daniel, Yang,  Yinghui Catherine and Zhang, Jie. (2015). ``Optimal Budget Allocation across Search Advertising Markets," \textit{INFORMS Journal on Computing}, 27(2): 285-300.

\bibitem[Yang et al.(2016)]{Yang2016} Yang, Yanwu, Yang, Yinghui Catherine, Liu, Dengpan and Zeng, Daniel. (2016). ``Dynamic Budget Allocation in Competitive Search Advertising," (June 1, 2016). Available at SSRN. \\ \url{https://ssrn.com/abstract=2912054}

\bibitem[Yang et al.(2018)]{Yang2018} Yang, Yanwu, Li, Xin, Zeng, Daniel, Jansen, Bernard, J. (2018), ``Aggregate effects of advertising decisions: a complex systems look at search engine advertising via an experimental study," \textit{Internet Research}, 28(4), 1079-1102.

\bibitem[Yao and Mela(2011)]{YaoMela2011} Yao, Song, and Carl F. Mela. (2011). ``A dynamic model of sponsored search advertising," \textit{Marketing Science}, 30(3), 447-468.

\bibitem[Ye et al.(2014)]{Ye2014}Ye, Shengqi, Goker Aydin, and Shanshan Hu. (2014). ``Sponsored search marketing: Dynamic pricing and advertising for an online retailer." \textit{Management Science} 61(6), 1255-1274.


\bibitem[Zhang and Feng(2011)]{ZhangFeng2011} Zhang, Xiaoquan, and Juan Feng. (2011). ``Cyclical bid adjustments in search-engine advertising," \textit{Management Science}, 57(9), 1703-1719.

\bibitem[Zhang et al.(2014)]{Zhang2014} Zhang, Jie, Yanwu Yang, Xin Li, Rui Qin, and Daniel Zeng. (2014). ``Dynamic dual adjustment of daily budgets and bids in sponsored search auctions." \textit{Decision support systems}, 57, 105-114. 















\end{thebibliography}
\end{document}